\documentclass[showpacs,twocolumn,aps,prb]{revtex4}
\usepackage{amsmath}
\usepackage{amsfonts}
\usepackage{amssymb}
\usepackage{amsthm}
\usepackage{graphicx}

\begin{document}

\title{Dielectric function of the semiconductor hole liquid:\\
Full frequency and wave vector dependence}
\author{John Schliemann}
\affiliation{Institute for Theoretical Physics, University of Regensburg,
D-93040 Regensburg, Germany}
\date{July 2011}

\begin{abstract}
We study the dielectric function of the homogeneous semiconductor hole liquid
of $p$-doped bulk III-V zinc-blende semiconductors within random phase
approximation. The single-particle physics of the hole system is modeled
by Luttinger's four-band Hamiltonian in its spherical approximation. 
Regarding the Coulomb-interacting hole liquid, the full
dependence of the zero-temperature 
dielectric function on wave vector and frequency is
explored. The imaginary part of the dielectric function is analytically
obtained in terms of complicated but fully elementary expressions, while
in the result for the real part nonelementary one-dimensional integrations
remain to be performed. The correctness of these two independent calculations 
is checked via Kramers-Kronig relations.

The mass difference between heavy and light holes, along with variations
in the background dielectric constant, leads to dramatic alternations in the
plasmon excitation pattern, and generically, two plasmon branches can be
identified. These findings are the result of the evaluation of the full
dielectric function and are not accessible via a high-frequency expansion.
In the static limit a beating of Friedel oscillations between the Fermi
wave numbers of heavy and light holes occurs.
\end{abstract}
\pacs{71.10.-w, 71.10.Ca, 71.45.Gm}
\maketitle

\section{Introduction}

The interacting electron gas, combined with a homogeneous neutralizing 
background, is one of the paradigmatic systems of many-body physics
\cite{Giuliani05,Mahan00,Bruus04}.
Albeit the result of drastic approximations,
its predictions provide a good description of important properties 
of three-dimensional bulk metals and, 
in the regime of lower carrier densities, 
$n$-doped semiconductors where the electrons reside in the s-type conduction 
band.

On the other hand, in a $p$-doped zinc-blende III-V semiconductor such as GaAs,
the defect electrons or holes occupy the p-type valence band whose more
complex band structure can be expected to significantly 
modify the electronic properties. Moreover, the most intensively studied
ferromagnetic semiconductors such as Mn-doped GaAs are in fact $p$-doped
with the holes playing the key role in the occurrence of carrier-mediated 
ferromagnetism among the localized Mn magnetic moments \cite{Jungwirth06}.
Thus, such $p$-doped bulk semiconductor systems lie at the very heart of the 
still growing field of spintronics \cite{Fabian07}, and therefore it appears 
highly desirable to gain a deeper understanding of their many-body physics.

{\em Ab-initio}-type approaches to the description of ferromagnetic
semiconductors constitute an important subfield of this endeavor,
and there is a lively discussion on strengths and weaknesses of
the pertaining concepts and numerical techniques\cite{Sato10}.
In the present paper we will follow a somewhat different route by developing
an analytical theory of the most prominent class of host materials
given by $p$-doped bulk III-V zinc-blende systems such as GaAs.
Specifically, we investigate the dielectric function 
of the interacting hole liquid within 
random phase approximation (RPA)\cite{Giuliani05,Mahan00,Bruus04}
where the non-interacting hole system in the valence band is described
by Luttinger's Hamiltonian in the spherical 
approximation\cite{Luttinger56}. We evaluate the zero-temperature
dielectric function
in the entire range of wave vectors and frequencies building upon a
recent study where the problem was analyzed in the static limit, and
in the case of large frequencies\cite{Schliemann10}. Another previous
work investigated, among other issues, properties of Hartree-Fock
solutions of the two-component carrier system consisting of heavy and light
holes\cite{Schliemann06}.
Moreover, very recently Kyrychenko and Ullrich have put forward a study
of holes in magnetically doped 
III-V systems~\cite{Kyrychenko09}
by modeling the band structure by an
$8\times 8$ $\vec k\cdot\vec p$ Hamiltonian (similar as in the 
present work) while disorder effects and interaction among the
carriers are treated by a combination of equations-of-motion techniques and
time-dependent density functional theory~\cite{Kyrychenko09,Kyrychenko11}.
Further below we will compare our fairly analytical results
with the ones of Ref.~\cite{Kyrychenko11} which rely heavier on
numerical evaluations.

Finally we mention a series of related recent studies of the dielectric
properties of two-dimensional fermionic systems 
(instead of three-dimensional bulk semiconductors) whose single-particle
states carry a non-trivial spinor structure. 
These include $n$-doped quantum wells with spin-orbit coupling
\cite{Pletyukhov06,Badalyan09,Agarwal11,Chesi11} 
and two-dimensional hole systems
\cite{Kernreiter10}. Other recent investigations have dealt with
planar graphene sheets where an effective spin is incorporated by the
sublattice degree of freedom\cite{Wunsch06,Stauber10,Pyatkovskiy09}.

The plan of the paper is as follows. In section \ref{preliminaries} we 
give an overview on elementary properties of the single-particle Hamiltonian
describing the band structure around the $\Gamma$-point, and on the
many-body formalism leading to the RPA result for the dielectric function. In
Section \ref{polar} we present explicit analytical expressions for the free
polarizability; the corresponding derivations are deferred to the appendices.
Section \ref{dielec} discusses physical properties of the dielectric function
and its full dependence on wave vector and frequency. Special attention is paid
to the static limit, and to the limit of large frequencies. We close with
conclusions and an outlook in section \ref{concl}.

\section{Preliminaries: Hamiltonian, eigensystem, and 
many-body formalism}
\label{preliminaries}

Luttinger's Hamiltonian describing heavy and light hole states around the
$\Gamma$-point in III-V zinc-blende semiconductors reads 
in its spherical approximation\cite{Luttinger56},
\begin{equation}
{\cal H}=\frac{1}{2m_{0}}\left(\left(\gamma_{1}+\frac{5}{2}\gamma_{2}\right)
\vec p^{2}-2\gamma_{2}\left(\vec p\cdot\vec S\right)^{2}\right)\,.
\label{Luttinger}
\end{equation}
Here $m_{0}$ is 
the bare electron mass, $\vec p$ is the hole lattice momentum, 
and $\vec S$ are spin-$3/2$-operators, resulting from
adding the $l=1$ orbital angular momentum to the $s=1/2$ electron spin.
The dimensionless Luttinger parameters $\gamma_{1}$ and $\gamma_{2}$ 
describe the valence
band of the specific material with effects of spin-orbit coupling being
included in $\gamma_{2}$. We note that , while the present work is mostly 
motivated by III-V semiconductors, the above model for the $\Gamma_{8}$ 
valence band also applies to other systems with zinc-blende or diamond 
structure including elemental semiconductors like Si and Ge, but also 
zero-gap semiconductors such as HgSe and HgTe.

The above Hamiltonian is rotationally invariant and 
commutes with the 
helicity operator $\lambda=(\vec k\cdot\vec S)/k$, where $\vec k=\vec p/ \hbar$ 
is the hole wave vector. The heavy (light) holes correspond to
$\lambda=\pm 3/2$ ($\lambda=\pm 1/2$) with the energy dispersions
\begin{equation}
\varepsilon_{h/l}(\vec k)=\frac{\hbar^{2}k^{2}}{2m_{h/l}}
\end{equation}
and heavy ($h$) and light ($l$) hole masses
$m_{h/l}=m_{0}/(\gamma_{1}\mp 2\gamma_{2})$.
The corresponding eigenstates are given by
\begin{equation}
\langle\vec r|\vec k,\lambda\rangle=
\frac{e^{i\vec k\vec r}}{\sqrt{V}}|\chi_{\lambda}(\vec k)\rangle\,,
\label{eigenstate}
\end{equation}
where $V$ is the volume of the system. Using the conventional basis of 
eigenstates of $S^{z}$ and introducing polar coordinates
$\vec k=k(\cos\varphi\sin\vartheta,\sin\varphi\sin\vartheta,\cos\vartheta)$ ,
the eigenspinors $|\chi_{\lambda}(\vec k)\rangle$ of the helicity operator
read explicitly~\cite{Schliemann06}
\begin{eqnarray}
|\chi_{\frac{3}{2}}(\vec k)\rangle & = & \left(
\begin{array}{c}
\cos^{3}\frac{\vartheta}{2}e^{-\frac{3i}{2}\varphi}\\
\sqrt{3}\cos^{2}\frac{\vartheta}{2}\sin\frac{\vartheta}{2}
e^{-\frac{i}{2}\varphi}\\
\sqrt{3}\cos\frac{\vartheta}{2}\sin^{2}\frac{\vartheta}{2}
e^{+\frac{i}{2}\varphi}\\
\sin^{3}\frac{\vartheta}{2}e^{+\frac{3i}{2}\varphi}
\end{array}
\right)\label{spinor1}\\
|\chi_{\frac{1}{2}}(\vec k)\rangle & = & \left(
\begin{array}{c}
-\sqrt{3}\cos^{2}\frac{\vartheta}{2}\sin\frac{\vartheta}{2}
e^{-\frac{3i}{2}\varphi}\\
\cos\frac{\vartheta}{2}\left(\cos^{2}\frac{\vartheta}{2}
-2\sin^{2}\frac{\vartheta}{2}\right)e^{-\frac{i}{2}\varphi}\\
\sin\frac{\vartheta}{2}\left(2\cos^{2}\frac{\vartheta}{2}
-\sin^{2}\frac{\vartheta}{2}\right)e^{+\frac{i}{2}\varphi}\\
\sqrt{3}\cos\frac{\vartheta}{2}\sin^{2}\frac{\vartheta}{2}
e^{+\frac{3i}{2}\varphi}
\end{array}
\right)
\label{spinor2}
\end{eqnarray}
and the remaining eigenspinors $|\chi_{-3/2}(\vec k)\rangle$, 
$|\chi_{-1/2}(\vec k)\rangle$ can be
obtained from the above ones by spatial inversion 
$\vartheta\mapsto\pi-\vartheta$, $\varphi\mapsto\varphi+\pi$.
In what follows,  mutual overlaps squared~\cite{Schliemann06}
between spinors will be of key importance,
\begin{eqnarray}
|\langle\chi_{\frac{3}{2}}(\vec k_{1})|\chi_{\frac{3}{2}}(\vec k_{2})\rangle|^{2} & = &
\left(\frac{1}{2}\left(1+\frac{\vec k_{1}\vec k_{2}}{k_{1}k_{2}}\right)\right)^{3}
\label{overlap1}\\
|\langle\chi_{\frac{1}{2}}(\vec k_{1})|\chi_{\frac{1}{2}}(\vec k_{2})\rangle|^{2} & = &
\frac{1}{8}\left(1+\frac{\vec k_{1}\vec k_{2}}{k_{1}k_{2}}\right)
\left(3\frac{\vec k_{1}\vec k_{2}}{k_{1}k_{2}}-1\right)^{2}
\label{overlap2}\\
|\langle\chi_{\frac{3}{2}}(\vec k_{1})|\chi_{\frac{1}{2}}(\vec k_{2})\rangle|^{2} & = &
\frac{3}{8}\left(1+\frac{\vec k_{1}\vec k_{2}}{k_{1}k_{2}}\right)^{2}
\left(1-\frac{\vec k_{1}\vec k_{2}}{k_{1}k_{2}}\right)
\label{overlap3}
\end{eqnarray}
Combining the above single-particle Hamiltonian with Coulomb repulsion among
holes and a neutralizing background, the dielectric function within RPA
at wave vector $\vec q$ and frequency $\omega$ 
is given by\cite{Giuliani05,Mahan00,Bruus04}
\begin{equation}
\varepsilon^{RPA}(\vec q,\omega)=1-V(\vec q)\chi_{0}(\vec q,\omega)\,. 
\end{equation}
Here $V(\vec q)$ is the 
Fourier transform of the interaction potential, and the
free polarizability reads
\begin{eqnarray}
 \chi_{0}(\vec q,\omega) 
& = & \frac{1}{(2\pi)^{3}}\sum_{\lambda_{1},\lambda_{2}}\int d^{3}k\Biggl[
\left|\langle\chi_{\lambda_{1}}(\vec k)|
\chi_{\lambda_{2}}(\vec k+\vec q)\rangle\right|^{2}\nonumber\\
 & & \cdot\frac{f(\vec k,\lambda_{1})-f(\vec k+\vec q,\lambda_{2})}
{\hbar\omega+i0-\left(\varepsilon_{\lambda_{2}}(\vec k+\vec q)
-\varepsilon_{\lambda_{1}}(\vec k)\right)}\Biggr]
\label{chi01}
\end{eqnarray}
with $f(\vec k,\lambda)$ being Fermi functions. In what follows we will
concentrate on the case of zero temperature and Coulomb repulsion,
$V(\vec q)=e^{2}/(\varepsilon_{r}\varepsilon_{0}q^{2})$ where
$\varepsilon_{r}$ is a background dielectric constant taking into
account screening by deeper bands.

\section{The free polarizability}
\label{polar}

We now present our analytical results for the real and imaginary part
of the free polarizability. Details of the derivations are can be found
in appendices \ref{appreal} and \ref{appimaginary}. A discussion of the
physical properties of the corresponding dielectric function follows
further below in section \ref{dielec}.
Defining
\begin{eqnarray}
 & & \chi_{hh}(\vec q,\omega)=\frac{1}{(2\pi)^{3}}\int_{k\leq k_{h}}d^{3}k
\left(\frac{1}{2}+\frac{3}{2}
\frac{\left(\vec k\cdot(\vec k+\vec q)\right)^{2}}
{k^{2}(\vec k+\vec q)^{2}}\right)\nonumber\\
 & & \qquad\qquad\Biggl[\frac{1}{\hbar\left(\omega+i0\right)
-\left(\varepsilon_{h}(\vec k+\vec q)-
\varepsilon_{h}(\vec k)\right)}\nonumber\\
 & & \qquad\qquad
-\frac{1}{\hbar\left(\omega+i0\right)+\left(\varepsilon_{h}(\vec k+\vec q)-
\varepsilon_{h}(\vec k)\right)}\Biggr]\,,
\label{chiHH}\\
 & & \chi_{hl}(\vec q,\omega)=\frac{1}{(2\pi)^{3}}\int_{k\leq k_{h}}d^{3}k
\left(\frac{3}{2}-\frac{3}{2}
\frac{\left(\vec k\cdot(\vec k+\vec q)\right)^{2}}
{k^{2}(\vec k+\vec q)^{2}}\right)\nonumber\\
 & & \qquad\qquad\Biggl[\frac{1}{\hbar\left(\omega+i0\right)
-\left(\varepsilon_{l}(\vec k+\vec q)-
\varepsilon_{h}(\vec k)\right)}\nonumber\\
 & & \qquad\qquad
-\frac{1}{\hbar\left(\omega+i0\right)+\left(\varepsilon_{l}(\vec k+\vec q)-
\varepsilon_{h}(\vec k)\right)}\Biggr]\,,\label{chiHL}
\end{eqnarray}
one can formulate the polarizability (\ref{chi01}) as follows,
\begin{equation}
\chi_{0}(\vec q,\omega)=\sum_{\alpha,\beta\in\{h,l\}}
\chi_{\alpha\beta}(\vec q,\omega)\,,
\label{chi02}
\end{equation}
where the remaining quantities $\chi_{ll}(\vec q,\omega)$ and
$\chi_{lh}(\vec q,\omega)$ are given by (\ref{chiHH}) and (\ref{chiHL})
via the replacement $h\leftrightarrow l$, and $k_{h}$ ($k_{l}$) is the Fermi
wave number for heavy (light) holes corresponding to the common
Fermi energy $\varepsilon_{f}$.\cite{notechi}
Introducing the obvious decomposition
$\chi_{\alpha\beta}(\vec q,\omega)=R_{\alpha\beta}(\vec q,\omega)
+iI_{\alpha\beta}(\vec q,\omega)$ with real functions
$R_{\alpha\beta}(\vec q,\omega)$, $I_{\alpha\beta}(\vec q,\omega)$ 
($\alpha,\beta\in\{h,l\}$), we will now analyze the real and imaginary part
of the free polarizability of the hole gas. The respective expressions
to be presented below are the result of independent calculations and
perfectly fulfil Kramers-Kronig relations\cite{Giuliani05,Mahan00,Bruus04}.

\subsection{The real part of the free polarizability}

Following the steps detailed in appendix \ref{appreal} the real part
of the free polarizability can be obtained as
\begin{widetext}
\begin{align}
 & R_{hh}(\vec q,\omega)+R_{hl}(\vec q,\omega)=\frac{-m_{h}}{(2\pi\hbar)^{2}}
\Biggl[2k_{h}+\frac{q}{4}
\frac{1}{(\varepsilon_{h}(q))^{2}}\Biggl(
\left(4\varepsilon_{f}\varepsilon_{h}(q)
-\left(\varepsilon_{h}(q)+\hbar\omega\right)^{2}\right)
\ln\left|\frac{\varepsilon_{h}(q)+\hbar\omega+\hbar^{2}qk_{h}/m_{h}}
{\varepsilon_{h}(q)+\hbar\omega-\hbar^{2}qk_{h}/m_{h}}\right| &
\nonumber\\
 & \qquad\qquad\qquad\qquad\qquad\qquad\qquad\qquad\qquad\qquad
+\left(4\varepsilon_{f}\varepsilon_{h}(q)
-\left(\varepsilon_{h}(q)-\hbar\omega\right)^{2}\right)
\ln\left|\frac{\varepsilon_{h}(q)-\hbar\omega+\hbar^{2}qk_{h}/m_{h}}
{\varepsilon_{h}(q)-\hbar\omega-\hbar^{2}qk_{h}/m_{h}}\right|
\Biggr)\Biggr] & \nonumber\\
 & \qquad\qquad\qquad
-\frac{3}{(2\pi\hbar)^{2}}\left(m_{h}-m_{l}\right)k_{h} & \nonumber\\
 & \qquad\qquad\qquad+\frac{3}{(2\pi\hbar)^{2}}\frac{m_{h}q}{32}
\left(1-\frac{m_{h}}{m_{l}}\right)^{2}
\left[\int_{0}^{2k_{h}/q}dy y
\ln\left|\frac{1-\hbar\omega/\varepsilon_{l}(q)+y+(1-m_{l}/m_{h})y^{2}/4}
{1-\hbar\omega/\varepsilon_{l}(q)-y+(1-m_{l}/m_{h})y^{2}/4}\right|
+\left(\omega\mapsto-\omega\right)\right] & \nonumber\\
 & \qquad\qquad\qquad+\frac{3}{(2\pi\hbar)^{2}}\frac{m_{h}q}{8}
\Biggl[\frac{\varepsilon_{h}(q)}{\hbar\omega}
\left(1+\frac{\hbar\omega}{\varepsilon_{h}(q)}\right)^{2}
{\cal P}\int_{0}^{2k_{h}/q}dy
\frac{y}{y^{2}+4\hbar\omega/\varepsilon_{h}(q)}
\Biggl(\ln\left|\frac{1-\hbar\omega/\varepsilon_{h}(q)+y}
{1-\hbar\omega/\varepsilon_{h}(q)-y}\right| & \nonumber\\
 & \qquad\qquad\qquad\qquad\qquad\qquad\qquad\qquad\qquad\qquad
-\ln\left|\frac{1-\hbar\omega/\varepsilon_{l}(q)+y+(1-m_{l}/m_{h})y^{2}/4}
{1-\hbar\omega/\varepsilon_{l}(q)-y+(1-m_{l}/m_{h})y^{2}/4}\right|\Biggr)
+\left(\omega\mapsto-\omega\right)\Biggr] & \nonumber\\
 & \qquad\qquad\qquad-\frac{3}{(2\pi\hbar)^{2}}\frac{m_{h}q}{8}
\left[\frac{\varepsilon_{h}(q)}{\hbar\omega}
\left(1-\frac{\hbar\omega}{\varepsilon_{h}(q)}\right)^{2}\int_{0}^{2k_{h}/q}dy
\frac{1}{y}\ln\left|\frac{1-\hbar\omega/\varepsilon_{h}(q)+y}
{1-\hbar\omega/\varepsilon_{h}(q)-y}\right|+\left(\omega\mapsto-\omega\right)
\right] & \nonumber\\
 & \quad+\frac{3}{(2\pi\hbar)^{2}}\frac{m_{h}q}{8}
\left[\frac{\varepsilon_{h}(q)}{\hbar\omega}
\left(1-\frac{\hbar\omega}{\varepsilon_{l}(q)}\right)^{2}\int_{0}^{2k_{h}/q}dy
\frac{1}{y}
\ln\left|\frac{1-\hbar\omega/\varepsilon_{l}(q)+y+(1-m_{l}/m_{h})y^{2}/4}
{1-\hbar\omega/\varepsilon_{l}(q)-y+(1-m_{l}/m_{h})y^{2}/4}\right|
+\left(\omega\mapsto-\omega\right)\right]\,, &
\label{real1}
\end{align}
\end{widetext}
where $(\omega\mapsto-\omega)$ denotes terms with the sign
of the frequency changed compared to the preceding expression, and 
the remaining contribution $R_{ll}(\vec q,\omega)+R_{lh}(\vec q,\omega)$
follows via $h\leftrightarrow l$. In the limit $m_{h}=m_{l}$
the first two lines in Eq.~(\ref{real1}) express the result for the
standard textbook case of a fermion gas without spin-orbit coupling
\cite{Giuliani05,Mahan00,Bruus04},
while all other terms vanish in this limit and represent corrections
arising from $m_{h}\neq m_{l}$. The contribution in the third line of
Eq.~(\ref{real1}) is constant, i.e. independent of $\vec q$ and $\omega$.
However, in the limit of large frequencies this term cancels against
the terms in the last two lines of the above equation such that
$\lim_{\omega\to\infty}\chi_{0}(\vec q,\omega)=0$. The integral occurring
in the fourth line of Eq.~(\ref{real1}) is elementary but lengthy 
(cf. appendix \ref{appreal}) while all other integrals cannot
be cast be cast into elementary expressions. Note that in the fifth line
of the above expression a proper Cauchy principal value
(denoted by $\cal{P}$) occurs. This mathematical detail arises from the 
Dirac identity, and the corresponding integral does for negative
frequency $\omega<0$ not converge in the general sense. The occurrence
of such nontrivial principal values is also a technical difference to
the standard jellium model.

\subsection{The imaginary part of the free polarizability}

\begin{figure}
\begin{center}
\includegraphics[width=8.5cm]{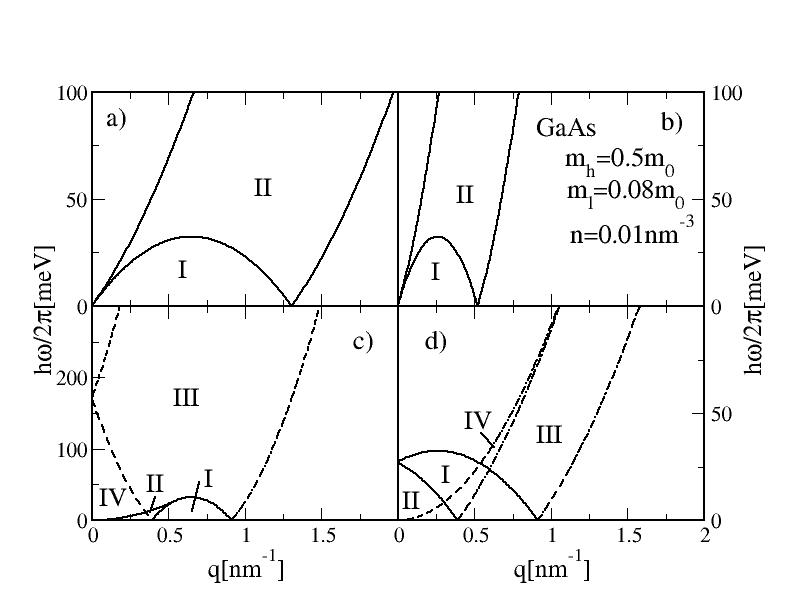}
\end{center}
\caption{Regions of nonvanishing contributions to a): $I_{hh}(\vec q,\omega)$,
b): $I_{ll}(\vec q,\omega)$, c): $I_{hl}^{+}(\vec q,\omega)$
(solid lines) and  $I_{hl}^{-}(\vec q,\omega)$ (dashed lines), 
and d): $I_{lh}^{+}(\vec q,\omega)$
(solid lines) and  $I_{lh}^{-}(\vec q,\omega)$ (dashed lines);
cf. tables~\ref{tableHH}-\ref{tableLH}.
We have chosen the mass parameters of GaAs, $m_{h}=0.5m_{0}$,
$m_{l}=0.08m_{0}$, and a hole density of $n=0.01{\rm nm}^{-3}$.
\label{fig1}}
\end{figure}

As the free polarizability $\chi_{0}(\vec r,t)$
is a real quantity, let us concentrate
on non-negative frequencies $\omega\geq 0$.
The regions of non-zero contribution $I_{hh}(\vec q,\omega)$ 
in the $q$-$\omega$-plane are given
explicitly in table~\ref{tableHH} and are depicted for typical system
parameters in Fig.~\ref{fig1}a). In region I and II 
$I_{hh}(\vec q,\omega)$ is given by
\begin{widetext}
\begin{eqnarray}
{\rm I:}\qquad
I_{hh}(\vec q,\omega) & = & \frac{-1}{4\pi q}\frac{m_{h}^{2}}{\hbar^{4}}
\hbar\omega\left[2-\frac{3}{8}
\left(1+\frac{\varepsilon_{h}(q)}{\hbar\omega}\right)^{2}
\ln\left|1+\frac{\hbar\omega}{\varepsilon_{f}}\right|
-\frac{3}{8}
\left(1-\frac{\varepsilon_{h}(q)}{\hbar\omega}\right)^{2}
\ln\left|1-\frac{\hbar\omega}{\varepsilon_{f}}\right|\right]\,,
\label{iHH1}\\
{\rm II:}\qquad
 I_{hh}(\vec q,\omega) & = & \frac{-1}{4\pi q}\frac{m_{h}^{2}}{\hbar^{4}}
\hbar\omega\Biggl[
\frac{2\varepsilon_{f}}{\hbar\omega}
-\left(1-\frac{\varepsilon_{h}(q)}{\hbar\omega}\right)^{2}
\left(\frac{\hbar\omega}{2\varepsilon_{h}(q)}
-\frac{3}{4}
\ln\left|\frac{\hbar^{2}qk_{h}/m_{h}}{\varepsilon_{h}(q)-\hbar\omega}\right|
\right)\nonumber\\
 & & \qquad\qquad\qquad
-\frac{3}{8}\left(1+\frac{\varepsilon_{h}(q)}{\hbar\omega}\right)^{2}
\ln\frac{4\varepsilon_{h}(q)(\varepsilon_{f}+\hbar\omega)}
{(\varepsilon_{h}(q)+\hbar\omega)^{2}}\Biggr]\,,
\label{iHH2}
\end{eqnarray}
\end{widetext}
respectively,
and are zero for all other values of $q$ and $\omega$. The region boundaries
given in  table~\ref{tableHH} are completely analogous to the ones
found for a standard jellium gas of spinless particles with mass $m_{h}$
and Fermi momentum $k_{h}$; for more details see appendix~\ref{appimaginary}.
The contributions to the imaginary part occurring in these regions are,
however, clearly different from the standard case.
The regions of nonvanishing contributions of and the corresponding expressions
for $I_{ll}(\vec q,\omega)$ can be obtained directly via the replacement
$h\mapsto l$.

The cases of the remaining expressions $I_{hl}(\vec q,\omega)$ and 
$I_{lh}(\vec q,\omega)$ are substantially more complicated. 
It is useful to distinguish two separate terms,
\begin{equation}
I_{hl}(\vec q,\omega)=I_{hl}^{+}(\vec q,\omega)
-I_{hl}^{-}(\vec q,\omega)\,,
\label{IHL0}
\end{equation}
and likewise for $I_{lh}(\vec q,\omega)$.
The corresponding
regions of nonzero contribution to $I_{hl}^{\pm}(\vec q,\omega)$
and $I_{lh}^{\pm}(\vec q,\omega)$
are given in tables~\ref{tableHL} and
\ref{tableLH}, respectively, and plotted in Figs.~\ref{fig1}c) and d) for
typical parameters. Now defining
\begin{widetext}
\begin{eqnarray}
G_{\pm}(q,\omega;k_{1},k_{2};m_{1},m_{2})
 & = & \frac{3}{8\pi q}\frac{m_{1}}{\hbar^{2}}
\Biggl[\pm\left(\frac{q^{2}}{2}\pm\frac{m_{2}\omega}{\hbar}\right)^{2}
\frac{\hbar}{2m_{1}\omega}
\left(\ln\frac{k_{1}}{k_{2}}-\frac{1}{2}
\ln\left|\frac{k_{1}^{2}\mp 2m_{1}\omega/ \hbar}
{k_{2}^{2}\mp 2m_{1}\omega/ \hbar}\right|\right)\nonumber\\
 & & \qquad+\left(q^{2}-\left(1-\frac{m_{2}}{m_{1}}\right)
\left(\frac{q^{2}}{2}\pm\frac{m_{2}\omega}{\hbar}\right)\right)
\frac{1}{2}
\ln\left|\frac{k_{1}^{2}\mp 2m_{1}\omega/ \hbar}
{k_{2}^{2}\mp 2m_{1}\omega/ \hbar}\right|\nonumber\\
 & & \qquad-\frac{1}{4}\left(1-\frac{m_{2}}{m_{1}}\right)^{2}
\left(\frac{1}{2}\left(k_{1}^{2}-k_{2}^{2}\right)
\pm\frac{m_{1}\omega}{\hbar}\ln\left|\frac{k_{1}^{2}\mp 2m_{1}\omega/ \hbar}
{k_{2}^{2}\mp 2m_{1}\omega/ \hbar}\right|\right)
\Biggr]
\end{eqnarray} 
\end{widetext}
$I_{hl}^{+}(\vec q,\omega)$ in regions I and II can be expressed as
\begin{eqnarray}
{\rm I:}\quad
I_{hl}^{+}(\vec q,\omega) & = & 
G_{+}(q,\omega;k_{h},\underline{k}_{h}^{+};m_{h},m_{l})\,,
\label{IHL1}\\
{\rm II:}\quad
I_{hl}^{+}(\vec q,\omega) & = & 
G_{+}(q,\omega;\overline{k}_{h}^{+},\underline{k}_{h}^{+};m_{h},m_{l})\,,
\label{IHL2}
\end{eqnarray}
respectively, where
\begin{eqnarray}
\underline{k}_{h}^{\pm} & = & \frac{q}{1-\frac{m_{l}}{m_{h}}}
\left|1-\sqrt{1-\left(1-\frac{m_{l}}{m_{h}}\right)
\left(1\pm\frac{\hbar\omega}{\varepsilon_{l}}\right)}\right|\,,
\label{lowHL}\\
\overline{k}_{h}^{\pm} & = & \frac{q}{1-\frac{m_{l}}{m_{h}}}
\left(1+\sqrt{1-\left(1-\frac{m_{l}}{m_{h}}\right)
\left(1\pm\frac{\hbar\omega}{\varepsilon_{l}}\right)}\right)\,.
\label{highHL}
\end{eqnarray}
The nonzero contributions to $I_{hl}^{-}(\vec q,\omega)$ in regions III and IV
of table~\ref{tableHL} are given by
\begin{eqnarray}
{\rm III:}\quad
I_{hl}^{-}(\vec q,\omega) & = & 
G_{-}(q,\omega;k_{h},\underline{k}_{h}^{-};m_{h},m_{l})\,,
\label{IHL3}\\
{\rm IV:}\quad
I_{hl}^{-}(\vec q,\omega) & = & 
G_{-}(q,\omega;\overline{k}_{h}^{-},\underline{k}_{h}^{-};m_{h},m_{l})\,,
\label{IHL4}
\end{eqnarray}

The nonvanishing contributions to $I_{lh}^{\pm}(\vec q,\omega)$
can be expressed in a similar manner. For $I_{lh}^{+}(\vec q,\omega)$
in regions I and II of table~\ref{tableLH} one finds
\begin{eqnarray}
{\rm I:}\quad
I_{lh}^{+}(\vec q,\omega) & = & 
G_{+}(q,\omega;k_{l},\underline{k}_{l}^{+};m_{l},m_{h})\,,
\label{ILH1}\\
{\rm II:}\quad
I_{lh}^{+}(\vec q,\omega) & = & 
G_{+}(q,\omega;\overline{k}_{l}^{+},\underline{k}_{l}^{+};m_{l},m_{h})\,,
\label{ILH2}
\end{eqnarray}
with
\begin{eqnarray}
\underline{k}_{l}^{\pm} & = & \frac{q}{\frac{m_{h}}{m_{l}}-1}
\left|1-\sqrt{1+\left(\frac{m_{h}}{m_{l}}-1\right)
\left(1\pm\frac{\hbar\omega}{\varepsilon_{h}}\right)}\right|\,,
\label{lowLH}\\
\overline{k}_{l}^{\pm} & = & \frac{q}{\frac{m_{h}}{m_{l}}-1}
\left(1+\sqrt{1+\left(\frac{m_{h}}{m_{l}}-1\right)
\left(1\pm\frac{\hbar\omega}{\varepsilon_{h}}\right)}\right)\,.
\label{highLH}
\end{eqnarray}
Likewise, the contributions to $I_{lh}^{-}(\vec q,\omega)$
in regions I and II are given by
\begin{eqnarray}
{\rm III:}\quad
I_{lh}^{-}(\vec q,\omega) & = & 
G_{-}(q,\omega;k_{l},\underline{k}_{l}^{-};m_{l},m_{h})\,,
\label{ILH3}\\
{\rm IV:}\quad
I_{lh}^{-}(\vec q,\omega) & = & 
G_{-}(q,\omega;\overline{k}_{l}^{-},\underline{k}_{l}^{-};m_{l},m_{h})\,,
\label{ILH4}
\end{eqnarray}

\begin{table}
\begin{tabular}{|c|c|}
\hline
 I & $q\leq 2k_{h}\,\wedge\,\hbar\omega\leq\hbar^{2}qk_{h}/m_{h}
-\varepsilon_{h}(q)$\\
\hline
 II & 
$[q\leq 2k_{h}\,\wedge\,\hbar^{2}qk_{h}/m_{h}-\varepsilon_{h}(q)
\leq\hbar\omega$\\
 & \hspace{1.5cm}$\leq
\hbar^{2}qk_{h}/m_{h}+\varepsilon_{h}(q)]$\\ 
 & $\vee\,\,[q\geq 2k_{h}\,\wedge\,-\hbar^{2}qk_{h}/m_{h}+\varepsilon_{h}(q)
\leq\hbar\omega$\\
 & \hspace{1.5cm}$\leq\hbar^{2}qk_{h}/m_{h}+\varepsilon_{h}(q)]$\\
\hline
\end{tabular}
\vspace{0.2cm}
\caption{Boundaries of regions of nonzero imaginary contribution
$I_{hh}(\vec q,\omega)$. The boundaries for $I_{ll}(\vec q,\omega)$ are
obtained via the replacement $h\mapsto l$.
\label{tableHH}}
\end{table}
\begin{table}
\begin{tabular}{|c|c|}
\hline
 I & $(1-\sqrt{m_{l}/m_{h}})k_{h}\leq q\leq(1+\sqrt{m_{l}/m_{h}})k_{h}$\\
 & $\wedge\,\hbar\omega\leq\hbar^{2}qk_{h}/m_{l}
-\varepsilon_{l}(q)-(m_{h}/m_{l}-1)\varepsilon_{f}$\\
\hline
II & $[q\leq(1-m_{l}/m_{h})k_{h}$\\
 & $\wedge\,\hbar^{2}qk_{h}/m_{l}
-\varepsilon_{l}(q)-(m_{h}/m_{l}-1)\varepsilon_{f}\leq\hbar\omega$\\
 & \hspace{0.8cm}$\leq\varepsilon_{h}(q)/(1-m_{l}/m_{h})]$\\
\hline
 III & 
$[q\leq(1-\sqrt{m_{l}/m_{h}})k_{h}$\\
& $\wedge\,-\hbar^{2}qk_{h}/m_{l}
+\varepsilon_{l}(q)+(m_{h}/m_{l}-1)\varepsilon_{f}\leq\hbar\omega$\\
 & \hspace{0.8cm}$\leq\hbar^{2}qk_{h}/m_{l}
+\varepsilon_{l}(q)+(m_{h}/m_{l}-1)\varepsilon_{f}]$\\
 & \\
 & $\vee\,\,[(1-\sqrt{m_{l}/m_{h}})k_{h}\leq q\leq(1+\sqrt{m_{l}/m_{h}})k_{h}$\\
& $\wedge\,\hbar\omega\leq\hbar^{2}qk_{h}/m_{l}
+\varepsilon_{l}(q)+(m_{h}/m_{l}-1)\varepsilon_{f}]$\\
 
 & \\
 & $\vee\,\,[q\geq(1+\sqrt{m_{l}/m_{h}})k_{h}$\\
& $\wedge\,-\hbar^{2}qk_{h}/m_{l}
+\varepsilon_{l}(q)+(m_{h}/m_{l}-1)\varepsilon_{f}\leq\hbar\omega$\\
 & \hspace{0.8cm}$\leq\hbar^{2}qk_{h}/m_{l}
+\varepsilon_{l}(q)+(m_{h}/m_{l}-1)\varepsilon_{f}]$\\
\hline
 IV & 
$q\leq(1-\sqrt{m_{l}/m_{h}})k_{h}$\\
& $\wedge\,\hbar\omega\leq-\hbar^{2}qk_{h}/m_{l}
+\varepsilon_{l}(q)+(m_{h}/m_{l}-1)\varepsilon_{f}$\\
\hline
\end{tabular}
\vspace{0.2cm}
\caption{Boundaries of regions of nonzero contributions to
$I_{hl}^{+}(\vec q,\omega)$ (regions I, II) and 
$I_{hl}^{-}(\vec q,\omega)$ (regions III, IV).
\label{tableHL}}
\end{table}
\begin{table}
\begin{tabular}{|c|c|}
\hline
 I & $[q\leq(\sqrt{m_{h}/m_{l}}+1)k_{l}$\\
& $\wedge\,-\hbar^{2}qk_{l}/m_{h}
-\varepsilon_{h}(q)-(m_{l}/m_{h}-1)\varepsilon_{f}\leq\hbar\omega$\\ 
& \hspace{0.8cm}$\leq\hbar^{2}qk_{l}/m_{h}
-\varepsilon_{h}(q)-(m_{l}/m_{h}-1)\varepsilon_{f}]$\\
\hline
 II & $q\leq(\sqrt{m_{h}/m_{l}}-1)k_{l}$\\
& $\wedge\,\hbar\omega\leq-\hbar^{2}qk_{l}/m_{h}
-\varepsilon_{h}(q)-(m_{l}/m_{h}-1)\varepsilon_{f}$\\
\hline  
 III & $[(\sqrt{m_{h}/m_{l}}-1)k_{l}\leq q\leq(\sqrt{m_{h}/m_{l}}+1)k_{l}$\\
 & $\wedge\,\hbar\omega\leq\hbar^{2}qk_{l}/m_{h}
+\varepsilon_{h}(q)+(m_{l}/m_{h}-1)\varepsilon_{f}]$\\
& \\
 & $\vee\,\,[q\geq(\sqrt{m_{h}/m_{l}}+1)k_{l}$\\
 & $\wedge\,-\hbar^{2}qk_{l}/m_{h}
+\varepsilon_{h}(q)+(m_{l}/m_{h}-1)\varepsilon_{f}\leq\hbar\omega$\\
 & \hspace{0.8cm}$\leq\hbar^{2}qk_{l}/m_{h}
+\varepsilon_{h}(q)+(m_{l}/m_{h}-1)\varepsilon_{f}]$\\
\hline
 IV & $[q\leq(m_{h}/m_{l}-1)k_{l}$\\
& $\wedge\,\hbar^{2}qk_{l}/m_{h}
+\varepsilon_{h}(q)+(m_{l}/m_{h}-1)\varepsilon_{f}\leq\hbar\omega$\\
 & \hspace{0.8cm}$\leq\varepsilon_{l}(q)/(m_{h}/m_{l}-1)]$\\
\hline
\end{tabular}
\vspace{0.2cm}
\caption{Boundaries of regions of nonzero contributions to
$I_{lh}^{+}(\vec q,\omega)$ (regions I, II) and 
$I_{lh}^{-}(\vec q,\omega)$ (regions III, IV).
\label{tableLH}}
\end{table}

\section{The dielectric function}
\label{dielec}

Let us now analyze the RPA dielectric function resulting from the
above free polarizability. We first concentrate on the effect of the mass
difference between heavy and light holes. To this end we eliminate effects
of the dielectric background by putting $\varepsilon_{r}=1$, and we fix
the total density $n=n_{h}+n_{l}$, $n_{h/l}=k_{h/l}^{3}/3\pi^{2}$, 
to $n=0.01{\rm nm}^{-3}$.
\begin{figure}
\begin{center}
\includegraphics[width=8.5cm]{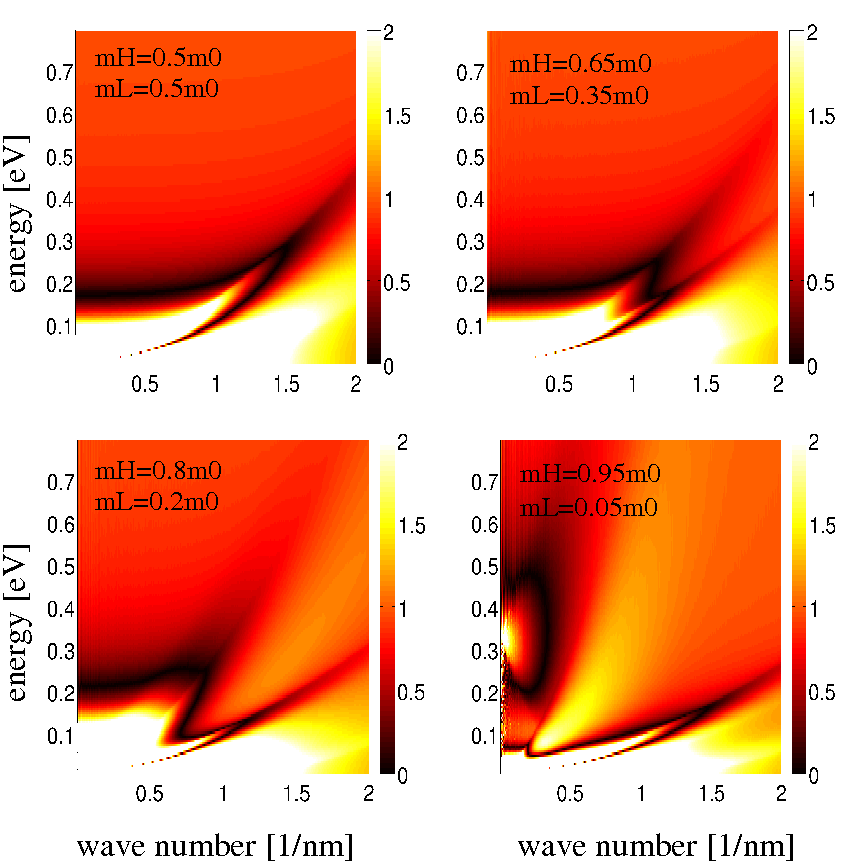}
\end{center}
\caption{The modulus $|{\rm Re}(\varepsilon^{RPA}(\vec q,\omega))|$ of the real 
part of the RPA dielectric function as a function of wave number $q$ and energy 
$\hbar\omega$ for a model system with $\varepsilon_{r}=1$. 
The ratio of heavy and light mass is varied at constant $m_{H}+m_{l}=m_{0}$,
and the total hole density is $n=0.01{\rm nm}^{-3}$.
\label{fig2}}
\end{figure}
\begin{figure}
\begin{center}
\includegraphics[width=8.5cm]{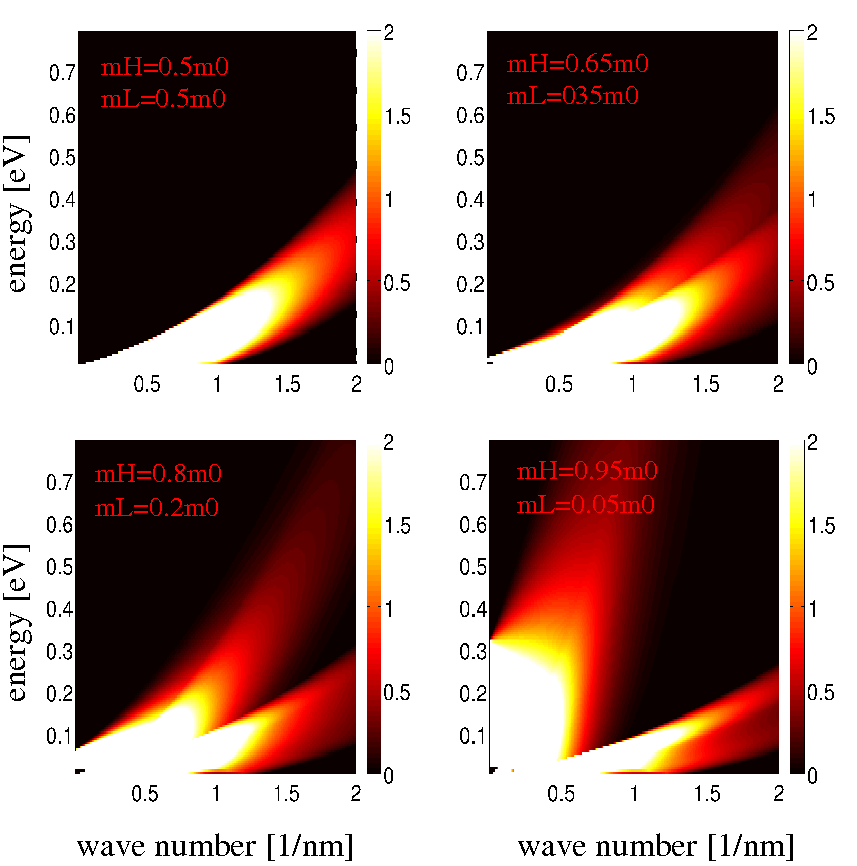}
\end{center}
\caption{The imaginary part ${\rm Im}(\varepsilon^{RPA}(\vec q,\omega))$ of the 
the RPA dielectric function as a function of wave number $q$ and energy 
$\hbar\omega$ for a model system with $\varepsilon_{r}=1$. 
The ratio of heavy and light mass is varied at constant $m_{H}+m_{l}=m_{0}$,
and the total hole density is $n=0.01{\rm nm}^{-3}$.
\label{fig3}}
\end{figure}
\begin{figure}
\begin{center}
\includegraphics[width=8.5cm]{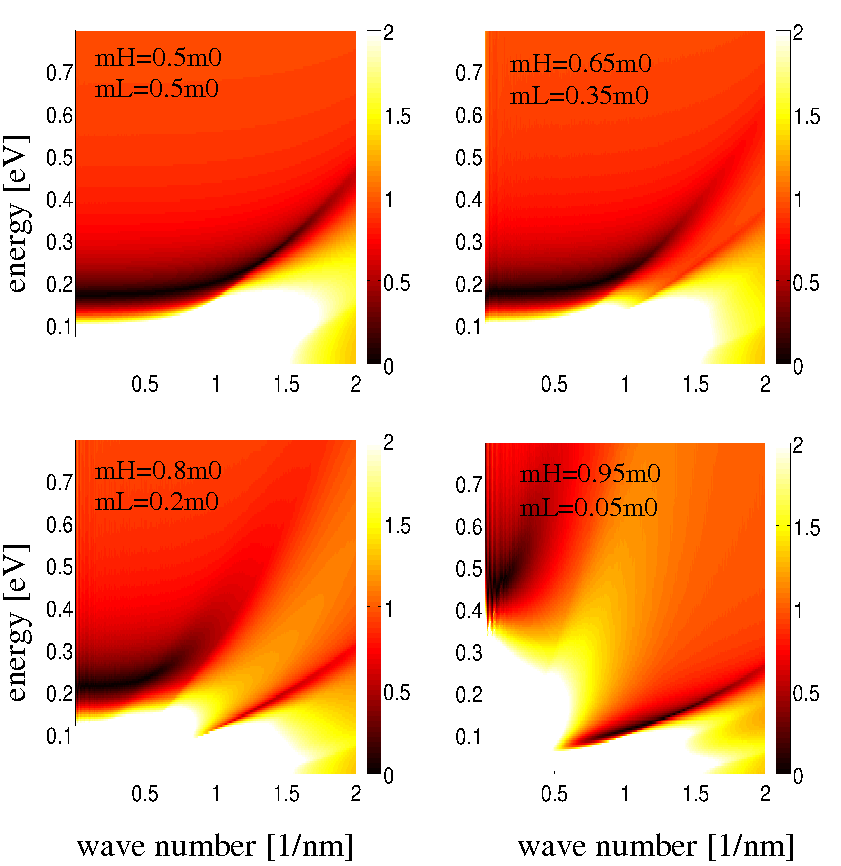}
\end{center}
\caption{The modulus $|\varepsilon^{RPA}(\vec q,\omega)|$ of
the RPA dielectric function as a function of wave number $q$ and energy 
$\hbar\omega$ for a model system with $\varepsilon_{r}=1$. 
The ratio of heavy and light mass is varied at constant $m_{H}+m_{l}=m_{0}$,
and the total hole density is $n=0.01{\rm nm}^{-3}$. The dark areas indicate 
zeros of the dielectric function corresponding to plasmon excitations.
\label{fig4}}
\end{figure}
Figs.~\ref{fig2}, \ref{fig3} show the real\cite{notereal} and imaginary part
of the dielectric function as a function of wave number and frequency
in a color-coded density plot, whereas in \ref{fig4} the modulus
of $\varepsilon^{RPA}(\vec q,\omega)$ is shown. The top left panel in each figure
illustrates the textbook case\cite{Giuliani05,Mahan00,Bruus04} 
of equal masses $m_{H}=m_{l}=m_{0}/2$ with its well-known plasmon dispersion
$\omega(q)$ determined by $\varepsilon^{RPA}(\vec q,\omega(q))=0$. With
increasing mass difference between heavy and light holes a more complex
structure arises and the plasmon dispersion splits into two branches
as seen in the bottom panels of Fig.~\ref{fig4}: A branch with comparatively
high energies at small wave numbers is accompanied by a branch at lower 
energies and large wave vectors. It is an interesting speculation whether
one can interpret these two plasmon branches in analogy to phonons: On one 
branch both heavy and light holes possibly perform (speaking in classical
terms) joint collective 
oscillations of charge density (analogous to acoustic phonons), while on the 
other branch they oscillate opposite to each other (similar to optical 
phonons). We leave this particular issue to future investigations.
\begin{figure}
\begin{center}
\includegraphics[width=8.5cm]{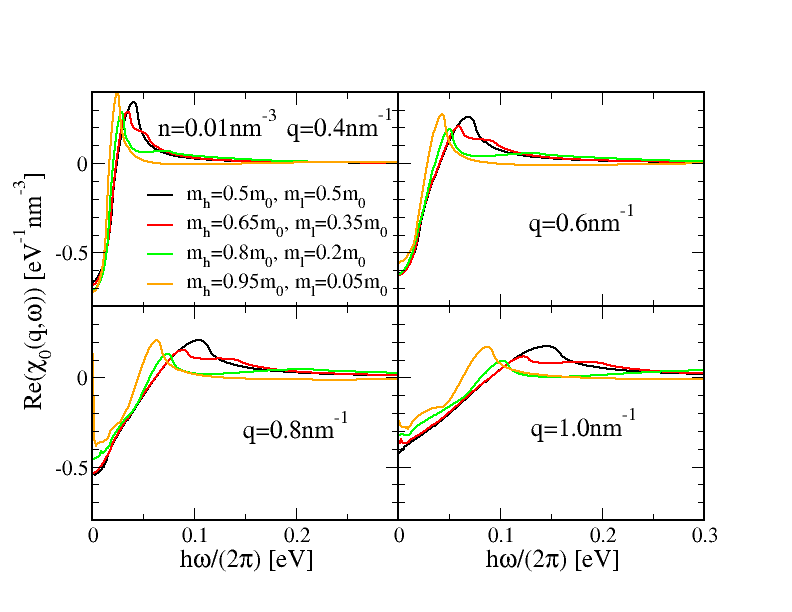}
\end{center}
\caption{The real part of the free polarizability $\chi_{0}(\vec q,\omega)$
as a function of frequency at different wave vectors 
for the same choice of heavy and light hole masses as in the previous figures.
\label{fig5}}
\end{figure}
\begin{figure}
\begin{center}
\includegraphics[width=8.5cm]{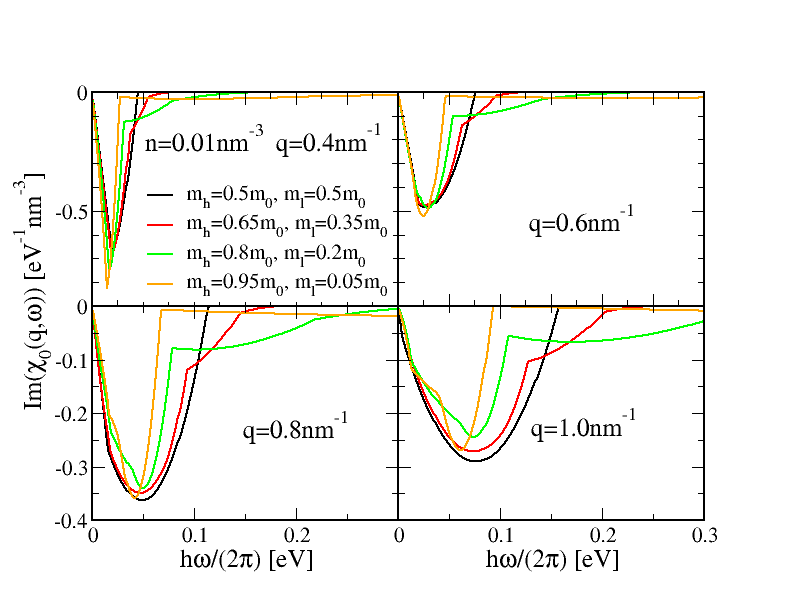}
\end{center}
\caption{The imaginary part of the free polarizability $\chi_{0}(\vec q,\omega)$
as a function of frequency at different wave vectors 
for the same choice of heavy and light hole masses as in the previous figures.
\label{fig6}}
\end{figure}

Finally, Figs.~\ref{fig5}, \ref{fig6} show the free polarizability
as a function of frequency at different wave vectors for the
same choice of heavy and light hole masses as in the previous figures.

\begin{figure}
\begin{center}
\includegraphics[width=8.5cm]{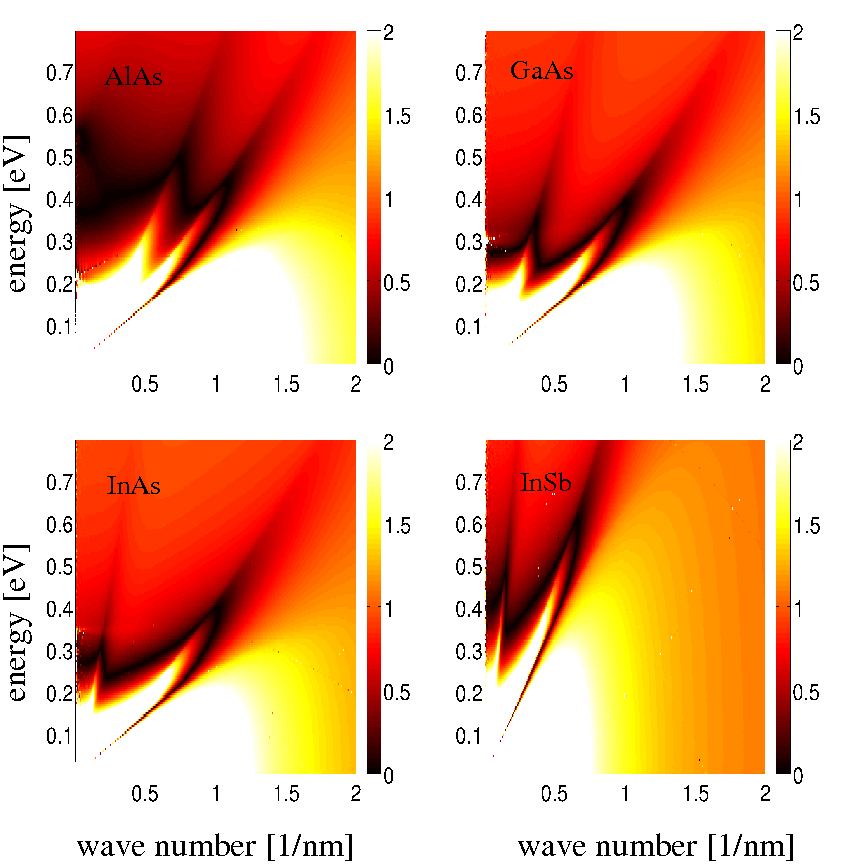}
\end{center}
\caption{The modulus $|{\rm Re}(\varepsilon^{RPA}(\vec q,\omega))|$ of the real 
part of the RPA dielectric function as a function of wave number $q$ and energy 
$\hbar\omega$ for various semiconductor systems at a total hole density 
of $n=0.35{\rm nm}^{-3}$.
\label{fig7}}
\end{figure}

\begin{figure}
\begin{center}
\includegraphics[width=8.5cm]{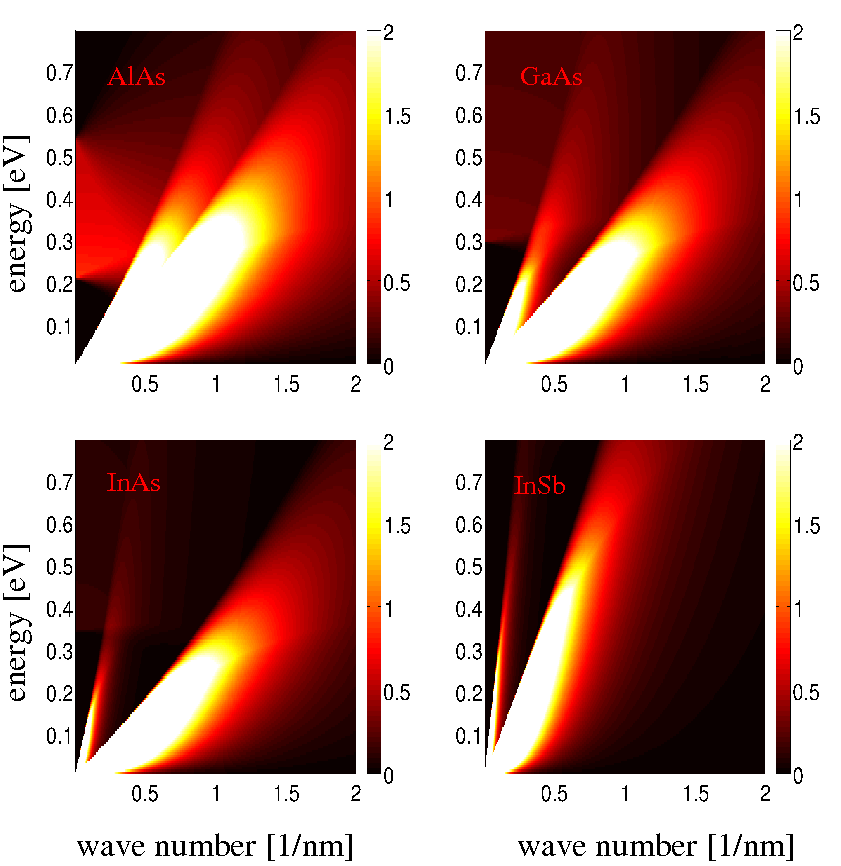}
\end{center}
\caption{The imaginary part ${\rm Im}(\varepsilon^{RPA}(\vec q,\omega))$ of the 
the RPA dielectric function as a function of wave number $q$ and energy 
$\hbar\omega$ for various semiconductor systems at a total hole density 
of $n=0.35{\rm nm}^{-3}$.
\label{fig8}}
\end{figure}

\begin{figure}
\begin{center}
\includegraphics[width=8.5cm]{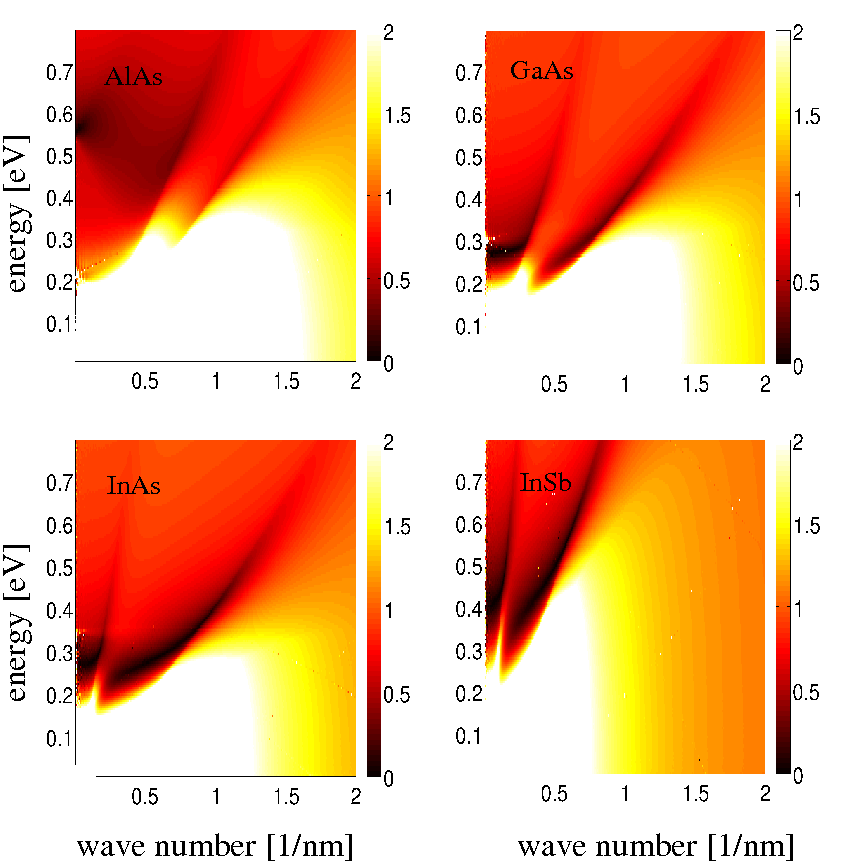}
\end{center}
\caption{The modulus $|\varepsilon^{RPA}(\vec q,\omega)|$ of
the RPA dielectric function as a function of wave number $q$ and energy 
$\hbar\omega$ for various semiconductor systems at a total hole density 
of $n=0.35{\rm nm}^{-3}$. The dark areas indicate zeros of the dielectric
function corresponding to plasmon excitations.
\label{fig9}}
\end{figure}

\begin{figure}
\begin{center}
\includegraphics[width=8.5cm]{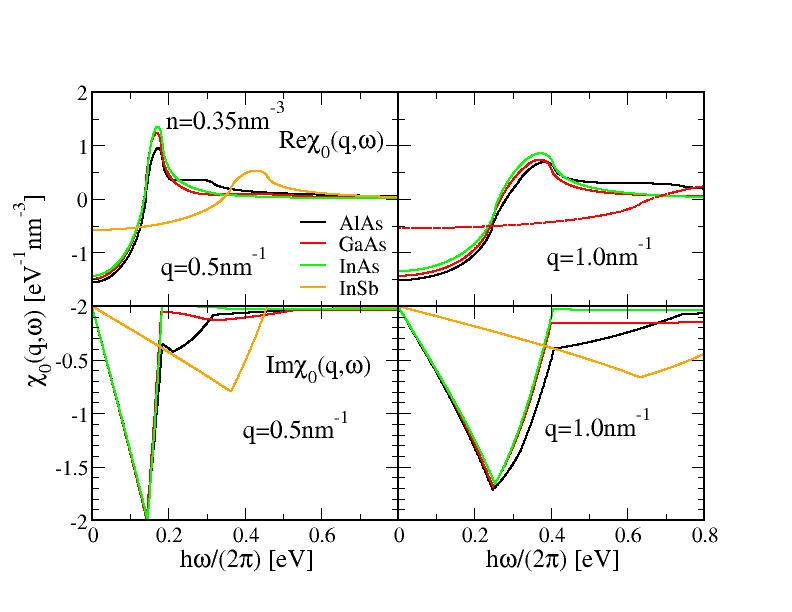}
\end{center}
\caption{The real (top panels) and imaginary (bottom panels)
part of the free polarizability $\chi_{0}(\vec q,\omega)$
as a function of frequency at different wave vectors 
for the same semiconductor systems as before (cf. table \ref{tablepara}).
\label{fig10}}  
\end{figure}

Let us now discuss our results for the dielectric function with respect
to concrete III-V semiconductors. In order to make contact to typical
ferromagnetic semiconductor systems\cite{Jungwirth06},
and to compare with results of
Ref.\cite{Kyrychenko11} we choose here a higher carrier density
of $n=0.35{\rm nm}^{-3}$. We consider four typical III-V systems whose
relevant parameters\cite{Vurgaftman01} are given in table~\ref{tablepara}.
Note that now also the background dielectric constant $\varepsilon_{r}$
plays a nontrivial role.
In Figs.~\ref{fig7}, \ref{fig8}, \ref{fig9} we have plotted the
real\cite{notereal} and imaginary part, and the modulus, respectively,
of the dielectric function as a function of wave number and frequency.
As seen from Fig.~\ref{fig9}, the zeros of the dielectric function
$\varepsilon^{RPA}(\vec q,\omega)$ defining the plasmon excitations 
form a clearly more complex pattern than in the standard jellium liquid, and,
as in the previous case, two dispersion branches can be identified.
In particular, the plasmon excitation in GaAs at small wave vector
occurs slightly below $0.3{\rm eV}$ which agrees very well with Fig.~4
of Ref.\cite{Kyrychenko11} where a more complex model for the band structure
was used. However, differently from the findings there, we can identify two
plasmon dispersion branches with small damping.
Moreover, In Fig.~\ref{fig10} we show the free polarizability 
$\chi_{0}(\vec q,\omega)$ as a function of frequency at different wave vectors 
for the same semiconductor systems. Again, the imaginary part for GaAs agrees 
nicely with data given in Fig.~2 of Ref.\cite{Kyrychenko11}. In this regime
the imaginary part of the free polarizability is dominated by transitions
between heavy-hole states, i.e. the main contributions is 
$I_{hh}(\vec q,\omega)$, in accordance with Ref.\cite{Kyrychenko11}.
\begin{table}
\begin{tabular}{c|c|c|c|}
  & $\frac{m_{h}}{m_{0}}$ & $\frac{m_{l}}{m_{0}}$ & $\varepsilon_{r}$ \\ 
\hline
AlAs & 0.47 & 0.18 & 10.0 \\
GaAs & 0.5 & 0.08 & 12.8  \\
InAs & 0.5 & 0.026 & 14.5 \\
InSb & 0.2 & 0.015 & 18.0 \\
\end{tabular}
\vspace{0.2cm}
\caption{Heavy and light hole masses long with background dielectric
constants for various III-V semiconductors\cite{Vurgaftman01}. 
\label{tablepara}}
\end{table}

\subsection{Static limit}

\begin{figure}
\begin{center}
\includegraphics[width=8.5cm]{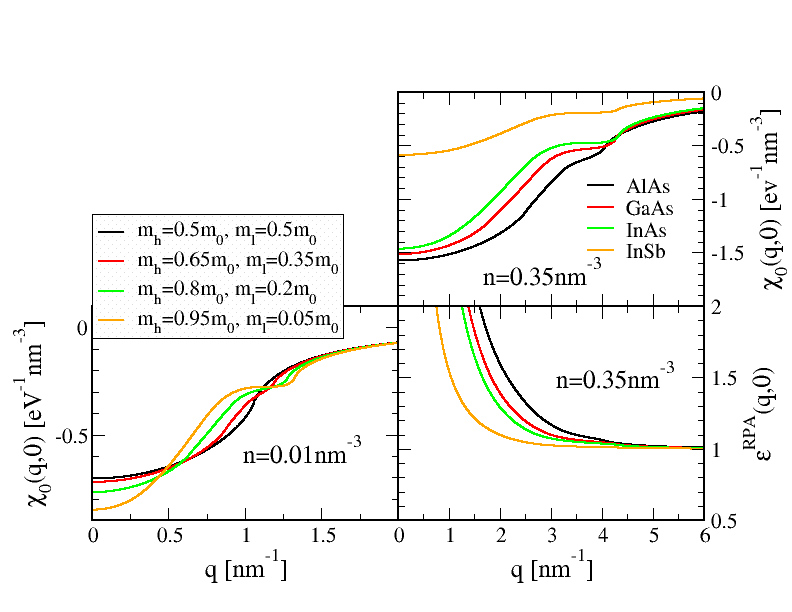}
\end{center}
\caption{Left panel: The static free polarizability $\chi_{0}(\vec q,0)$
for the same choice of parameters as in Figs.~\ref{fig2}-\ref{fig6}.
Right panels: $\chi_{0}(\vec q,0)$ and $\varepsilon^{RPA}(\vec q,0)$
the same III-V semiconductor systems as in Figs.~\ref{fig7}-\ref{fig10}
(cf. table \ref{tablepara}).
\label{fig11}}
\end{figure}

In the static limit $\omega=0$, the rather complex contributions
(\ref{real1}) to the free polarizability of the hole system simplify
considerably to~\cite{Schliemann10,notecorr1}
\begin{eqnarray}
\chi_{0}(\vec q,0) & = & -\frac{m_{h}}{\pi^{2}\hbar^{2}}k_{h}
\left(1+3\left(\frac{q}{2k_{h}}\right)^{2}\right)
L\left(\frac{q}{2k_{h}}\right)\nonumber\\
 & & -\frac{m_{l}}{\pi^{2}\hbar^{2}}k_{l}
\left(1+3\left(\frac{q}{2k_{l}}\right)^{2}\right)
L\left(\frac{q}{2k_{l}}\right)\nonumber\\
 & & +\frac{3\left(\sqrt{m_{h}}+\sqrt{m_{l}}\right)^{2}}{4\pi^{2}\hbar^{2}}
\frac{q^{2}}{k_{h}+k_{l}}
L\left(\frac{q}{k_{h}+k_{l}}\right)\nonumber\\
 & & -\frac{3\left(m_{h}-m_{l}\right)}{4\pi^{2}\hbar^{2}}
\left(k_{h}-k_{l}\right)\left(1-L\left(\frac{q}{k_{h}+k_{l}}\right)\right)
\nonumber\\
 & & +\frac{3m_{h}}{2\pi^{2}\hbar^{2}}qH\left(\frac{q}{2k_{h}}\right)
+\frac{3m_{l}}{2\pi^{2}\hbar^{2}}qH\left(\frac{q}{2k_{l}}\right)\nonumber\\
 & & -\frac{3\left(m_{h}+m_{l}\right)}{2\pi^{2}\hbar^{2}}
qH\left(\frac{q}{k_{h}+k_{l}}\right)\,,
\label{static}
\end{eqnarray}
where $L(x)$ is the so-called Lindhard correction,
\begin{equation}
L(x)
=\frac{1}{2}+\frac{1-x^{2}}{4x}\ln\left|\frac{1+x}{1-x}\right|\,,
\label{Lindhard}
\end{equation}
and the function $H$ is defined as
\begin{eqnarray}
H(x) & = & \frac{1}{2}\int_{0}^{1/x}dy\frac{1}{y}\ln\left|\frac{1+y}{1-y}\right|
\nonumber\\
 & = & \left\{
\begin{array}{ll}
\frac{\pi^{2}}{4}-\sum_{n=0}^{\infty}\frac{x^{2n+1}}{(2n+1)^{2}} & |x|\leq 1 \\
\sum_{n=0}^{\infty}\frac{\left(\frac{1}{x}\right)^{2n+1}}{(2n+1)^{2}} & |x|\geq 1
\end{array}
\right.\,.
\label{H}
\end{eqnarray}
Details of the derivation of the above result can be found in 
appendix~\ref{appstatic}. 
Note that the static polarizability can entirely be expressed in terms of the 
arguments $k/2k_{h}$, $k/2k_{l}$, and $k/k_{h}+k_{l}$ with the latter one being
the harmonic mean of the two former. In the case $m_{h}=m_{l}$
(i.e. $k_{h}=k_{l}=:k_{F}$) one obtains the usual result
$\chi_{0}(\vec q,0)=-D(\varepsilon_{F})L(q/2k_{F})$ for charge
carriers without spin-orbit coupling where $D(\varepsilon)$
is the density of states \cite{Simion05}. For $m_{h}\neq m_{l}$,
however, the static polarizability (\ref{static}) has a clearly more 
complicated structure. Fig.~\ref{fig11} displays the static free
polarizability and dielectric function for the systems discussed
above. In particular, the data in the left panel at fixed
$m_{h}+m_{l}=m_{0}$ shows that the static polarizability develops
richer features with increasing difference in heavy and light hole mass.

Moreover, in the long-wave approximation 
$\chi_{0}(\vec q,0)\approx\chi_{0}(0,0)$
one recovers the usual Thomas-Fermi (TF) screening,
\begin{equation}
\varepsilon^{RPA}(\vec q,0)\approx 1-q^{2}_{TF}/q^{2}
\end{equation}
with a Thomas-Fermi wave number
$q^{2}_{TF}=(e^{2}/\varepsilon_{r}\varepsilon_{0})3n/(2\varepsilon_{f})$.

As discussed in Ref.~\cite{Schliemann10}, the full screened potential of
a pointlike probe charge $Q$,
\begin{equation}
\Phi(\vec r)=\frac{1}{(2\pi)^{3}}\int d^{3}q
\frac{\frac{Q}{\varepsilon_{r}\varepsilon_{0}q^{2}}}
{\varepsilon^{RPA}(\vec q)}e^{i\vec q\vec r}
\label{fourier}
\end{equation}
can conveniently be approximated using Lighthill's theorem~\cite{Lighthill58}
as
\begin{equation}
\Phi(r)\approx\frac{m_{h}}{m_{0}}\phi_{\infty}(2k_{h},r)
+\frac{m_{l}}{m_{0}}\phi_{\infty}(2k_{l},r)
\label{friedel}
\end{equation}
where
\begin{equation}
\phi_{\infty}(q,r)=\frac{Q}{4\pi\varepsilon_{0}a_{0}}\frac{2}{\pi}
\frac{1}{\left(\varepsilon_{r}\varepsilon^{RPA}(q)\right)^{2}}
\frac{\cos(qr)}{(qr)^{3}}
\end{equation}
and $a_{0}=4\pi\varepsilon_{0}\hbar^{2}/(m_{0}e^{2})$ 
being the usual Bohr radius. As a result, 
a beating of Friedel oscillations between the two 
wave numbers $2k_{h/l}$ (but not $k=(k_{h}+k_{l})/2$) takes 
place\cite{Schliemann10}. This beating is a peculiarity of the holes residing 
in the p-type valence band
and should be observable via similar scanning tunneling microscopy techniques
as used in metals \cite{Crommie93} and $n$-doped semiconductors
\cite{Kanisawa01}. Moreover, as theoretical studies have revealed, such 
oscillations can have a profound impact on the magnetic properties of
ferromagnetic semiconductors by giving way to the possibility of
noncollinear magnetic ordering\cite{Schliemann02,Fiete05}.

\subsection{Limit of large frequencies}

In the regime of large frequencies and small wave vectors.
one can expand the denominators in Eq.~(\ref{chi01})
assuming $\hbar\omega>>\varepsilon_{h/l}(\vec q)$ and 
$\hbar\omega>>(\hbar k_{h/l}/m_{h/l})\hbar q$. The result within the two 
leading orders reads\cite{Schliemann10,notecorr2}
\begin{eqnarray}
 &  & \varepsilon^{RPA}(\vec q,\omega)=1-\frac{1}{\omega^{2}}
\frac{e^{2}}{\varepsilon_{r}\varepsilon_{0}}\frac{1}{6\pi^{2}}
\left(\frac{1}{m_{h}}+\frac{1}{m_{l}}\right)
\left(k_{h}^{3}+k_{l}^{3}\right)\nonumber\\
 & & \qquad-\frac{1}{\omega^{4}}\frac{e^{2}\hbar^{2}}
{\varepsilon_{r}\varepsilon_{0}\pi^{2}}
\frac{1}{2}\left(\frac{1}{m_{h}^{3}}+\frac{1}{m_{l}^{3}}\right)
\Biggl[\frac{1}{5}q^{2}\left(k_{h}^{5}+k_{l}^{5}\right)\nonumber\\
 & & \qquad\qquad+\frac{1}{12}q^{4}\left(k_{h}^{3}+k_{l}^{3}\right)\Biggr]
\nonumber\\
& & \qquad-\frac{1}{\omega^{4}}\frac{e^{2}\hbar^{2}}
{\varepsilon_{r}\varepsilon_{0}\pi^{2}}
\Biggl[-\frac{1}{56}\left(\frac{1}{m_{h}}-\frac{1}{m_{l}}\right)^{3}
\left(k_{h}^{7}-k_{l}^{7}\right)\nonumber\\
 & & \qquad\qquad+\frac{21}{200}q^{2}
\left(\frac{1}{m_{h}^{3}}-\frac{1}{m_{l}^{3}}\right)
\left(k_{h}^{5}-k_{l}^{5}\right)\nonumber\\
 & & \qquad\qquad-\frac{3}{40}q^{2}\left(\frac{1}{m_{h}}-\frac{1}{m_{l}}\right)
\left(\frac{k_{h}^{5}}{m_{l}^{2}}-\frac{k_{l}^{5}}{m_{h}^{2}}\right)\Biggr]
\label{high}
\end{eqnarray}
For $m_{h}=m_{l}$ the first three lines of the above expression reproduce again
the standard textbook result \cite{Mahan00} while all other terms vanish in 
this limit. On the other hand, if $m_{h}\neq m_{l}$, contributions
in order $1/\omega^{4}$ occur that are {\em independent of the wave vector} 
$\vec q$. Such terms are absent in the case of the standard electron gas where
the contributions of order $1/\omega^{2n}$ are at least of order
$q^{2n-2}$ in the wave vector \cite{Mahan00}. The technical reason
why such contributions are present for the hole gas is that the expression
$\varepsilon_{\lambda_{2}}(\vec k+\vec q)-\varepsilon_{\lambda_{1}}(\vec k)$
in Eq.(\ref{chi01}) contains for $|\lambda_{1}|\neq|\lambda_{2}|$ an additive 
term which is independent of $k$ (and vanishes for $m_{h}=m_{l}$).
As a consequence, although the result~(\ref{high}) is the valid high frequency 
expansion of the dielectric function, it is not possible to obtain from it
a reliable expression for the plasmon dispersion  $\omega(q)$ defined by
$\varepsilon^{RPA}(\vec q,\omega(q))=0$. This statement holds
even for the long-wavelength plasma frequency $\omega(q=0)$
and is due to the fact that in any order in $(1/\omega^{2})$
the prefactor in the expansion contains contributions being of low order
(including zeroth order) in $q$.
As an example, relying on the expansion (\ref{high}) being of up to quartic
order in $1/\omega$, the condition $\varepsilon^{RPA}(\vec q,\omega(q))=0$
translates to\cite{Schliemann10,notecorr2}
\begin{eqnarray}
  & & \omega^{2}(q)=\left(\omega_{p}^{(0)}\right)^{2}
\Biggl[\frac{1}{2}+\frac{1}{2}\Bigl[1+4\Bigl(u\left(n^{1/3}a_{0}\right)
\nonumber\\
 & & \qquad\qquad+\left(v+w\right)
\frac{(qa_{0})^{2}}{n^{1/3}a_{0}}\Bigr)\Bigr]^{1/2}
\Biggr]+{\cal O}\left(q^{4}\right)\,.
\label{plasma1}
\end{eqnarray}
Here we have defined
\begin{equation}
\left(\omega_{p}^{(0)}\right)^{2}
=\frac{e^{2}}{\varepsilon_{r}\varepsilon_{0}}\frac{n}{2}
\left(\frac{1}{m_{h}}+\frac{1}{m_{l}}\right)\,,
\label{plasma2}
\end{equation}
and the density-independent coefficients $u$, $v$, and $w$ are given by
\begin{eqnarray}
 & & u=\frac{-Q\left(m_{h},m_{l}\right)}
{\left(3\pi^{2}\right)^{1/3}\left(m_{h}^{3/2}+m_{l}^{3/2}\right)^{2/3}}\nonumber\\
 & & \quad\times
\frac{3}{14}\left(\frac{1}{m_{h}}-\frac{1}{m_{l}}\right)^{3}
\left(m_{h}^{7/2}-m_{l}^{7/2}\right)\,,
\label{u}
\end{eqnarray}
\begin{equation}
v=\frac{2Q\left(m_{h},m_{l}\right)}{5\pi^{2}}
\left(\frac{1}{m_{h}^{3}}+\frac{1}{m_{l}^{3}}\right)
\left(m_{h}^{5/2}+m_{l}^{5/2}\right)\,,
\label{v}
\end{equation}
\begin{eqnarray}
 & & w=\frac{3Q\left(m_{h},m_{l}\right)}{10\pi^{2}}\Biggl[\frac{7}{5}
\left(\frac{1}{m_{h}^{3}}-\frac{1}{m_{l}^{3}}\right)
\left(m_{h}^{5/2}-m_{l}^{5/2}\right)\nonumber\\
 & & \qquad\qquad-
\left(\frac{1}{m_{h}}-\frac{1}{m_{l}}\right)
\left(\frac{m_{h}^{5/2}}{m_{l}^{2}}-\frac{m_{l}^{5/2}}{m_{h}^{2}}\right)\Biggr]
\label{w}
\end{eqnarray}
with the common prefactor
\begin{equation}
Q\left(m_{h},m_{l}\right)=\frac{\frac{\varepsilon_{r}}{4\pi}m_{0}}
{\left(\frac{1}{m_{h}}+\frac{1}{m_{l}}\right)^{2}}
\frac{(3\pi^{2})^{5/3}}{\left(m_{h}^{3/2}+m_{l}^{3/2}\right)^{5/3}}\,.
\end{equation}
Clearly, the coefficients $u$ and $w$ vanish for $m_{h}=m_{l}$ while
from $v$ one recovers usual textbook result for an electron gas without
spin-orbit coupling \cite{Mahan00}. However, if $m_{h}$ and $m_{l}$ differ
substantially, the neglected contributions occurring in higher order in the 
inverse frequency but being independent of or of low order in the wave vector
can substantially modify the plasmon excitations. This can even affect the
plasma frequency $\omega(q=0)$ at zero wave vector: E.g. for the parameters
of GaAs (cf. table~\ref{tablepara}) and a total hole density of 
$n=0.35{\rm nm}^{-3}$ one obtains from Eq.~(\ref{plasma1}) 
$\hbar\omega(q=0)\approx 0.8{\rm eV}$, in contrast to the value 
of slightly less than $0.3{\rm eV}$ found from a full evaluation of
the dielectric function (cf. Fig.~\ref{fig9} top right panel) which is also
in accordance with Ref.\cite{Kyrychenko11}. In summary, although the
expansion~\ref{high} is the correct description of the dielectric function
in the limit of large frequencies, it does not lead to reliable expressions
for the plasmon dispersion. This is due to peculiarities of the expansion
occurring for $m_{h}\neq m_{l}$. On the other hand, as seen from
e.g. Fig.~\ref{fig9}, the interplay between the hole mass difference and the 
background dielectric constant leads to plasmon excitations patterns
which differ dramatically from the textbook case of the jellium model.
This alternations, however, are not accurately described by expressions of the
type (\ref{plasma1}), in contrast to an earlier approach where results
for the full wave vector and frequency dependence of the dielectric function
were not available yet\cite{Schliemann10}.

\section{Conclusions and outlook}
\label{concl}

We have investigated the RPA dielectric function of the homogeneous 
semiconductor hole liquid in $p$-doped bulk III-V zinc-blende semiconductors.
The single-particle physics of the hole system is modeled
by Luttinger's four-band Hamiltonian in its spherical approximation. 
Regarding the Coulomb-interacting hole liquid, the full
dependence of the zero-temperature 
dielectric function on wave vector and frequency has been
explored. The imaginary part of the dielectric function is analytically
obtained in terms of complicated but fully elementary expressions, while
in the result for the real part nonelementary one-dimensional integrations
remain to be performed. The correctness of these two independent calculations 
is checked via Kramers-Kronig relations.

The mass difference between heavy and light holes, along with variations
in the background dielectric constant, leads to dramatic alternations in the
plasmon excitation pattern, and generically, two plasmon branches can be
identified. These findings are the result of the evaluation of the full
dielectric function and are not accessible via a high-frequency expansion.
In the static limit a beating of Friedel oscillations between the Fermi
wave numbers of heavy and light holes occurs.

Regarding future developments, possible extensions of the present work
could include the implementation of more general single-particle Hamiltonians
modeling the band structure. For instance, one could drop the spherical
approximation to the Luttinger Hamiltonian and consider 
parameters $\gamma_{2}\neq\gamma_{3}$. However, such a reduction of the
full rotational invariance to tetrahedral symmetry might prohibit
analytical progress as achieved here. However, we do not expect drastic
effects from such a generalization, in particular not since for the
generic material GaAs $\gamma_{2}$, $\gamma_{3}$ are very close to each
other\cite{Vurgaftman01}. Moreover, our results obtained here
for the spherically symmetric $4\times 4$ valence band Hamiltonian
agree, where overlapping, very reasonably with findings in 
Ref.~\cite{Kyrychenko11} where a more complicated $8\times 8$ band structure
model was evaluated numerically.

Having in mind ferromagnetic semiconductors such as Mn-doped GaAs, another
obvious extension would be a coupling to the hole spins by a homogeneous
Zeeman-type field mimicking the magnetization of the Mn ions. A technical
difficulty here lies in the fact that the resulting single-particle
Hamiltonian cannot be diagonalized any more in a convenient analytical fashion.
However, analytical progress might still be possible if the Zeeman coupling is
treated as a perturbation. This strategy leads of course to also consider
the spin susceptibility, i.e. spin-spin response function. For a standard
jellium systems of spin-$1/2$ fermions without spin-orbit coupling, this 
quantity is, up to constant prefactors, identical to the free electrical
polarization\cite{Giuliani05}. For the $4\times 4$ hole system studied here,
however, this simple relation is rendered invalid by the larger spin length
and, more importantly, the presence of manifest spin-orbit coupling. Thus,
a study of the spin susceptibility in a similarly analytical fashion
as done here for the electric polarizability and the dielectric function
appears also to be useful.

Finally, from the point of view of general many-particle physics, the inclusion
of so-called local many-body field factors would be an important step
towards correlations beyond RPA\cite{Giuliani05}. 
The practical treatment
of such local field factors in the presence of strong spin-orbit 
coupling, however, is still in its infancy.

\acknowledgments{This work was supported by DFG via SFB 689  
``Spin Phenomena in Reduced Dimensions''.}

\appendix

\begin{widetext}
\section{Calculation of the free polarizability}

\subsection{The real part} 
\label{appreal}

With the help of the Dirac identity
\begin{equation}
\frac{1}{x+i0}={\cal P}\frac{1}{x}-i\pi\delta(x)
\label{dirac}
\end{equation}
the contributions to the real part of the free polarizability of the hole
gas read
\begin{eqnarray}
R_{hh}(\vec q,\omega) & = & \frac{-1}{(2\pi\hbar)^{2}}\frac{m_{h}}{q^{2}}
{\cal P}\int_{0}^{k_{h}}dk k^{2}\int_{-1}^{1}dx
\left(1+3\frac{k^{2}+2kqx+q^{2}x^{2}}{k^{2}+2kqx+q^{2}}\right)\nonumber\\
 & & \qquad\qquad\qquad\times
\left[\frac{1}{1+(2k/q)x-2m_{h}\omega/(\hbar q^{2})}
+\left(\omega\mapsto-\omega\right)\right]\,,\\
R_{hl}(\vec q,\omega) & = & \frac{-1}{(2\pi\hbar)^{2}}\frac{m_{l}}{q^{2}}
{\cal P}\int_{0}^{k_{h}}dk k^{2}\int_{-1}^{1}dx
\left(3-3\frac{k^{2}+2kqx+q^{2}x^{2}}{k^{2}+2kqx+q^{2}}\right)\nonumber\\
 & & \qquad\qquad\qquad\times
\left[\frac{1}{1+(2k/q)x+(1-m_{l}/m_{h})k^{2}/q^{2}-2m_{l}\omega/(\hbar q^{2})}
+\left(\omega\mapsto-\omega\right)\right]\,.
\end{eqnarray}
The integration over the polar variable $x$ can be performed by applying the
identity
\begin{eqnarray}
\frac{k^{2}+2kqx+q^{2}x^{2}}{k^{2}+2kqx+q^{2}}\frac{1}{1+(2k/q)x+\alpha} & = & 
\left(\frac{q}{2k}\right)^{2}
-\frac{q/(8k)}{1-(q/k)^{2}\alpha}\left(1-\frac{q^{2}}{k^{2}}\right)
\frac{1}{x+q/(2k)+k/(2q)}\nonumber\\
 & & +\frac{q/(2k)}{1-(q/k)^{2}\alpha}\left(1-\frac{q^{2}}{2k^{2}}
(1+\alpha)\right)^{2}
\frac{1}{x+q(1+\alpha)/(2k)}
\end{eqnarray}
for $\alpha=-2m_{h}\omega/(\hbar q^{2})$ and 
$\alpha=(1-m_{l}/m_{h})k^{2}/q^{2}-2m_{l}\omega/(\hbar q^{2})$. Adding both
contributions, and introducing a dimensionless radial integration 
variable $y=2k/q$, the result can be formulated as
\begin{eqnarray}
R_{hh}(\vec q,\omega)+R_{hl}(\vec q,\omega) & = &\frac{-1}{(2\pi\hbar)^{2}}
\frac{q}{4}{\cal P}\int_{0}^{2k_{h}/q}dy y^{2}
\Biggl[\frac{m_{h}}{2}\frac{1}{y}
\ln\left|\frac{1-2m_{h}\omega/(\hbar q^{2})+y}
{1-2m_{h}\omega/(\hbar q^{2})-y}\right|+\frac{3}{y^{2}}(m_{h}-m_{l})\nonumber\\
& & +\frac{3m_{l}}{2}\frac{1}{y}
\ln\left|\frac{1-2m_{l}\omega/(\hbar q^{2})+y+(1-m_{l}/m_{h})y^{2}/4}
{1-2m_{l}\omega/(\hbar q^{2})-y+(1-m_{l}/m_{h})y^{2}/4}\right|\nonumber\\
& & +\frac{3m_{h}}{2}\frac{y}{y^{2}+8m_{h}\omega/(\hbar q^{2})}
\left(1-\frac{2}{y^{2}}\left(1-\frac{2m_{h}\omega}{\hbar q^{2}}\right)\right)^{2}
\ln\left|\frac{1-2m_{h}\omega/(\hbar q^{2})+y}
{1-2m_{h}\omega/(\hbar q^{2})-y}\right|\nonumber\\
& & -\frac{3m_{h}}{2}\frac{y}{y^{2}+8m_{h}\omega/(\hbar q^{2})}
\left(\frac{1}{2}\left(1+\frac{m_{l}}{m_{h}}\right)
-\frac{2}{y^{2}}\left(1-\frac{2m_{l}\omega}{\hbar q^{2}}\right)\right)^{2}
\nonumber\\
& & \qquad\qquad
\times\ln\left|\frac{1-2m_{l}\omega/(\hbar q^{2})+y+(1-m_{l}/m_{h})y^{2}/4}
{1-2m_{l}\omega/(\hbar q^{2})-y+(1-m_{l}/m_{h})y^{2}/4}\right|
+\left(\omega\mapsto-\omega\right)\Biggr]
\label{appr1}
\end{eqnarray}
Now, by rearranging the terms in the integrand and performing elementary
integrations, one obtains the result
(\ref{real1}). We note that also the integral in the third line of
Eq.~(\ref{real1}) is elementary but lengthy,
\begin{eqnarray}
 \int_{0}^{x}dy\,y\ln\left|\frac{ay^{2}+by+c}{ay^{2}-by+c}\right| & = & 
\frac{1}{2}\left(x^{2}-\frac{b^{2}}{2a^{2}}+\frac{c}{a}\right)
\ln\left|\frac{ay^{2}+by+c}{ay^{2}-by+c}\right|+\frac{b}{a}x\nonumber\\
& - &  \left\{
\begin{array}{ll}
\frac{b}{a}\sqrt{\frac{b^{2}}{4a^{2}}-\frac{c}{a}}
\left(
\tanh^{-1}\left(\frac{x+b/(2a)}{\sqrt{b^{2}/(4a^{2})-c/a}}\right)
+\tanh^{-1}\left(\frac{x-b/(2a)}{\sqrt{b^{2}/(4a^{2})-c/a}}\right)\right) &
\frac{b^{2}}{4a^{2}}-\frac{c}{a}\geq 0 \nonumber \\
\frac{b}{a}\sqrt{\frac{c}{a}-\frac{b^{2}}{4a^{2}}}
\left(
\tan^{-1}\left(\frac{x+b/(2a)}{\sqrt{c/a-b^{2}/(4a^{2})}}\right)
+\tan^{-1}\left(\frac{x-b/(2a)}{\sqrt{c/a-b^{2}/(4a^{2})}}\right)\right) &
\frac{b^{2}}{4a^{2}}-\frac{c}{a}\leq 0 
\end{array}
\right.\,,
\end{eqnarray}
whereas all other integrals in Eq.~(\ref{real1}) cannot be expressed via
elementary functions.

\subsubsection{Static limit}
\label{appstatic}

In the static limit $\omega\to 0$ one obtains from Eq.~(\ref{real1})
(or, alternatively, Eq.~(\ref{appr1}))
\begin{eqnarray}
R_{hh}(\vec q,0)+R_{hl}(\vec q,0) & = &
\frac{-m_{h}}{(2\pi\hbar)^{2}}
\left(2k_{h}+\frac{q/2}{\varepsilon_{h}(q)}
\left(\varepsilon_{f}-\varepsilon_{h}(q)\right)
\ln\left|\frac{\varepsilon_{h}(q)+\hbar^{2}qk_{h}/m_{h}}
{\varepsilon_{h}(q)-\hbar^{2}qk_{h}/m_{h}}\right|\right)
-\frac{3}{(2\pi\hbar)^{2}}\left(m_{h}-m_{l}\right)k_{h}\nonumber\\
 & + & \frac{3}{(2\pi\hbar)^{2}}\frac{m_{h}q}{16}
\left(1-\frac{m_{h}}{m_{l}}\right)^{2}
\sum_{\mu=\pm}y_{\mu}^{2}\int_{0}^{2k_{h}/(qy_{\mu})}dy y
\ln\left|\frac{1+y}{1-y}\right|\nonumber\\
 & + & \frac{3}{(2\pi\hbar)^{2}}m_{h}q\int_{0}^{2k_{h}/q}dy \frac{1}{y}
\ln\left|\frac{1+y}{1-y}\right|-\frac{3/2}{(2\pi\hbar)^{2}}
\left(m_{h}+m_{l}\right)\sum_{\mu=\pm}\int_{0}^{2k_{h}/(qy_{\mu})}dy \frac{1}{y}
\ln\left|\frac{1+y}{1-y}\right|\nonumber\\
 & - & \frac{3}{(2\pi\hbar)^{2}}m_{h}q\int_{0}^{2k_{h}/q}dy \frac{1}{y^{3}}
\left(\ln\left|\frac{1+y}{1-y}\right|
-\ln\left|\frac{1+y/y_{+}}{1-y/y_{+}}\right|
-\ln\left|\frac{1+y/y_{-}}{1-y/y_{-}}\right|\right)\,,
\label{appr2}
\end{eqnarray}
where we have split a part of the logarithmic terms in the integrand by
introducing $y_{\pm}:=2/(1\pm\sqrt{m_{l}/m_{h}})$. In order to simplify the
above expression we first note that the first term of the r.h.s can be 
written as $-m_{h}L(q/(2k_{h}))/(\pi\hbar)^{2}$ using the Lindhard
correction~(\ref{Lindhard}). Regarding the terms in the second and
third line involving a summation over $\mu=\pm$, the interchange
$h\leftrightarrow l$ leads to 
$y_{\pm}\leftrightarrow \bar y_{\pm}:=2/(1\pm\sqrt{m_{h}/m_{l}})$
fulfilling $2k_{l}/(q\bar y_{\pm})=\pm2k_{h}/(qy_{\pm})$ and
$m_{h}\bar y_{\pm}^{2}=m_{l}y_{\pm}^{2}$. From these observations it is easy to 
see that the terms with $\mu=-$ cancel against corresponding expressions
in $R_{ll}(\vec q,0)+R_{lh}(\vec q,0)$, and only the terms
with $\mu=+$ (being invariant under $h\leftrightarrow l$) contribute
to $\chi_{0}(\vec q,0)$. The first of these contributions
(second line in Eq.~(\ref{appr2})) can be expressed again via the
Lindhard correction, while the integrals in the third line involve the 
function $H(x)$ defined in Eq.~(\ref{H}). Finally, the last line
of Eq.~(\ref{appr2}) can be evaluated as
\begin{eqnarray}
 & & m_{h}\int_{0}^{2k_{h}/q}dy \frac{1}{y^{3}}
\left(\ln\left|\frac{1+y}{1-y}\right|
-\ln\left|\frac{1+y/y_{+}}{1-y/y_{+}}\right|
-\ln\left|\frac{1+y/y_{-}}{1-y/y_{-}}\right|\right)\nonumber\\
 & = & 2m_{h}\frac{q}{2k_{h}}L\left(\frac{q}{2k_{h}}\right)
-\frac{\left(\sqrt{m_{h}}+\sqrt{m_{l}}\right)^{2}}{2}
\frac{q}{k_{h}+k_{l}}L\left(\frac{q}{k_{h}+k_{l}}\right)
-\frac{\left(\sqrt{m_{h}}-\sqrt{m_{l}}\right)^{2}}{2}
\frac{q}{k_{h}-k_{l}}L\left(\frac{q}{k_{h}-k_{l}}\right)\,,
\end{eqnarray}
where the last term on the r.h.s. is odd under $h\leftrightarrow l$
and cancels agianst an analogous contribution in 
$R_{ll}(\vec q,0)+R_{lh}(\vec q,0)$. Now summing all  expressions one
ends up with the result (\ref{static}) for the free polarizability
$\chi_{0}(\vec q,0)$.

\subsection{The imaginary part} 
\label{appimaginary}

\subsubsection{$I_{hh}(\vec q,\omega)$ and $I_{ll}(\vec q,\omega)$}

Using the Dirac identity (\ref{dirac}) and performing the angular
integrations, $I_{hh}(\vec q,\omega)$ can be expressed as
\begin{eqnarray}
I_{hh}(\vec q,\omega) & = & \frac{-1}{4\pi q}\frac{m_{h}}{\hbar^{2}}
\int_{0}^{k_{h}}dk\Biggl[\left(2k+\frac{3}{2k}
\frac{(q^{2}/2+m_{h}\omega/ \hbar)^{2}-q^{2}k^{2}}{k^{2}-2m_{h}\omega/ \hbar}
\right)
\Theta\left(k-\left|\frac{q}{2}+\frac{m_{h}\omega}{\hbar q}\right|\right)
\nonumber\\
 & & \qquad\qquad\qquad-\left(2k+\frac{3}{2k}
\frac{(q^{2}/2-m_{h}\omega/ \hbar)^{2}-q^{2}k^{2}}{k^{2}+2m_{h}\omega/ \hbar}
\right)
\Theta\left(k-\left|\frac{q}{2}-\frac{m_{h}\omega}{\hbar q}\right|\right)
\Biggr]\,,
\label{appiHH1}
\end{eqnarray}
\end{widetext}
where $\Theta(x)$ denotes the Heaviside step function. Obviuosly, the 
step functions occurring in the above expression define the lower integration 
bound, and the pertaining discussion parallels the arguments appropriate
for the standard textbook case of a spinless Jellium model
\cite{Giuliani05,Mahan00,Bruus04}. However, for the sake of completeness,
and in order to point out important differences regarding the
remaining quantities $I_{hl}(\vec q,\omega)$ and $I_{lh}(\vec q,\omega)$ to
be analyzed below, let us briefly go into some details. Since
$I_{hh}(\vec q,-\omega)=-I_{hh}(\vec q,\omega)$ it is sufficient to concentrate
on $\omega\geq 0$. Then, if the first step function in (\ref{appiHH1})
leads to a non-zero contribution (i.e. has a positive
argument for some $k\in[0,k_{h}]$), this holds also for the second
step function. Thus, a necessary and sufficient condition for both
step function to contribute is $k_{h}-|q/2+m_{h}\omega/(\hbar q)|\geq 0$, which 
is equivalent to
\begin{equation}
\hbar q\frac{\hbar k_{h}}{m_{h}}-\varepsilon_{h}(q)\geq \hbar\omega
\label{ineqHH1}
\end{equation}
and can for non-negative frequencies only be fulfilled for
$q\leq 2k_{h}$. The last two inequalities define region I in 
table~\ref{tableHH}, and the corresponding expression (\ref{iHH1}) is
obtained by elementary integration. 

Let us now turn to the case where only the second step function contributes,
i.e. $k_{h}-|q/2-m_{h}\omega/(\hbar q)|\geq 0$ while 
inequality (\ref{ineqHH1}) is violated. Assuming 
$\hbar\omega\geq\varepsilon_{h}(q)$ we arrive at the condition
\begin{equation}
\hbar q\frac{\hbar k_{h}}{m_{h}}+\varepsilon_{h}(q)\geq \hbar\omega
\geq\hbar q\frac{\hbar k_{h}}{m_{h}}-\varepsilon_{h}(q)\,,
\label{ineqHH2}
\end{equation}
while the opposite assumption, $\hbar\omega\leq\varepsilon_{h}(q)$, leads to 
\begin{equation}
 \hbar\omega\geq-\hbar q\frac{\hbar k_{h}}{m_{h}}+\varepsilon_{h}(q)\,.
\end{equation}
The latter inequality is only a nontrivial condition if its r.h.s. is 
non-negative which is equivalent to $q\geq 2k_{h}$. In summary, the
second step function in expression (\ref{appiHH1}) contributes while the
first one yields zero if, and only if, (i) $q\leq 2k_{h}$ and 
inequality (\ref{ineqHH2}) is fulfilled, or (ii) $q\geq 2k_{h}$ and
\begin{equation}
\hbar q\frac{\hbar k_{h}}{m_{h}}+\varepsilon_{h}(q)\geq \hbar\omega
\geq-\hbar q\frac{\hbar k_{h}}{m_{h}}+\varepsilon_{h}(q)\,.
\label{ineqHH3}
\end{equation}
The above conditions define region II in table~\ref{tableHH}, and
and the corresponding expression (\ref{iHH2}) follows
again from elementary integration.

The remaining quantity $I_{ll}(\vec q,\omega)$ is obtained from the above
results via the replacement $h\mapsto l$.

\subsubsection{$I_{hl}(\vec q,\omega)$}

After performing the angular integrations, $I_{hl}(\vec q,\omega)$ can be
formulated in the form (\ref{IHL0}) with
\begin{widetext}
\begin{eqnarray}
I_{hl}^{\pm}(\vec q,\omega) & = & \frac{3}{8\pi q}\frac{m_{h}}{\hbar^{2}}
\int_{0}^{k_{h}}dk\Biggl[\frac{1/k}{k^{2}\mp 2m_{h}\omega/ \hbar}
\Biggl(
-\left(\frac{q^{2}}{2}\pm\frac{m_{l}\omega}{\hbar}\right)^{2}
+k^{2}\left(q^{2}-\left(1-\frac{m_{l}}{m_{h}}\right)
\left(\frac{q^{2}}{2}\pm\frac{m_{l}\omega}{\hbar}\right)\right)\nonumber\\
& & \qquad\qquad\qquad\qquad
-\frac{k^{4}}{4}\left(1-\frac{m_{l}}{m_{h}}\right)^{2}
\Biggr)
\Theta\left(k-\left|\frac{q}{2}+\left(1-\frac{m_{l}}{m_{h}}\right)
\frac{k^{2}}{2q}\pm\frac{m_{l}\omega}{\hbar q}\right|\right)\Biggr]\,.
\label{appiHL1}
\end{eqnarray}
\end{widetext}
We now have to discuss under which circumstances the step functions in the 
above expression lead to nonzero contributions. This is more complicated
than in the previous case since the arguments of the step functions depend
quadratically (and not only linearly) on the integration variable $k$.
The condition
\begin{equation}
\Theta\left(k-\left|\frac{q}{2}+\left(1-\frac{m_{l}}{m_{h}}\right)
\frac{k^{2}}{2q}\pm\frac{m_{l}\omega}{\hbar q}\right|\right)=1
\end{equation}
is equivalent to
\begin{equation}
(k+a)^{2}\geq b_{\pm}\quad\wedge\quad(k-a)^{2}\leq b_{\pm}
\label{ineqab}
\end{equation}
where we have defined 
\begin{eqnarray}
a & = & \frac{q}{1-m_{l}/m_{h}}\,,\\
b_{\pm} & = & a^{2}-\frac{2q}{1-m_{l}/m_{h}}
\left(\frac{q}{2}\pm\frac{m_{l}\omega}{\hbar q}\right)\nonumber\\
 & = & \frac{(m_{l}/m_{h})q^{2}}{(1-m_{l}/m_{h})^{2}}
\mp\frac{2m_{l}\omega/ \hbar}{1-m_{l}/m_{h}}\,.
\end{eqnarray}
Here and in what follows the upper (lower) sign refers always to 
$I_{hl}^{+}(\vec q,\omega)$ ($I_{hl}^{-}(\vec q,\omega)$).
Note that the step function in
$I_{hl}^{+}(\vec q,\omega)$ can, for
non-negative frequencies, only be nonzero if this is also
the case for $I_{hl}^{-}(\vec q,\omega)$.

Since $m_{l}<m_{h}$ we clearly have $a\geq 0$, and the second inequality
in (\ref{ineqab}) requires $b_{\pm}\geq 0$ which leads for the upper sign
to the condition
\begin{equation}
\hbar\omega\leq\frac{1}{1-m_{l}/m_{h}}\varepsilon_{h}(q)\,.
\label{ineqb}
\end{equation}
Moreover, an elementary discussion of the inequalities (\ref{ineqab}) yields
the following lower and upper boundaries for the integral (\ref{appiHL1})
after resolving the step function,
\begin{eqnarray}
\underline{l}_{\pm} & = & \min\left\{|a-\sqrt{b_{\pm}}|,k_{h}\right\}\,,\\
\overline{l}_{\pm} & = & \min\left\{a+\sqrt{b_{\pm}},k_{h}\right\}\,,
\end{eqnarray}
with $0\leq\underline{l}_{\pm}\leq\overline{l}_{\pm}\leq k_{h}$.
Nonzero contributions occur only for $\underline{l}_{\pm}< k_{h}$.
For the upper sign, the condition $\underline{l}_{\pm}\geq k_{h}$ implies
\begin{equation}
\hbar q\frac{\hbar k_{h}}{m_{l}}-\varepsilon_{l}(q)
-\left(\frac{m_{h}}{m_{l}}-1\right)\varepsilon_{f}\leq \hbar\omega
\label{ineqHL1}
\end{equation}
which is, for non-negative $\omega$, a nontrivial statement on if  
\begin{equation}
\left(1-\sqrt{m_{l}/m_{h}}\right)k_{h}\leq q\leq
\left(1+\sqrt{m_{l}/m_{h}}\right)k_{h}\,.
\label{ineqHL2}
\end{equation}
Conversely, inequality (\ref{ineqHL1}) implies $\underline{l}_{\pm}\geq k_{h}$
only for $a\leq k_{h}$, while in the case $a\geq k_{h}$, i.e.
\begin{equation}
q\leq\left(1-m_{l}/m_{h}\right)k_{h}\,,
\label{ineqHL3}
\end{equation}
it follows $k_{h}\leq a+\sqrt{b_{+}}$, and the inequality (\ref{ineqb})
limits the region of nonzero $I_{hl}^{+}(\vec q,\omega)$. 
The inequalities (\ref{ineqHL3}), (\ref{ineqHL1}), and (\ref{ineqb})
define region II in table~\ref{tableHL} with the integration
bounds being $\underline{l}_{+}=|a-\sqrt{b_{+}}|=:\underline{k}_{h}^{+}$
and $\overline{l}_{+}=a+\sqrt{b_{+}}=:\overline{k}_{h}^{+}$
as defined in Eqs.~(\ref{lowHL}),(\ref{highHL}).
On the other hand, inequality (\ref{ineqHL2}) along with the negation
of (\ref{ineqHL1}),
\begin{equation}
\hbar q\frac{\hbar k_{h}}{m_{l}}-\varepsilon_{l}(q)
-\left(\frac{m_{h}}{m_{l}}-1\right)\varepsilon_{f}\geq \hbar\omega\,,
\label{ineqHL4}
\end{equation}
define region I in table~\ref{tableHL}. Here again
$\underline{l}_{+}=\underline{k}_{h}^{+}$, and inequality
(\ref{ineqHL4}) further implies  $\overline{l}_{+}=k_{h}$.
Note that the condition (\ref{ineqb}) is always fulfilled
if (\ref{ineqHL1}) is valid since
\begin{eqnarray}
 & &\hbar q\frac{\hbar k_{h}}{m_{l}}-\varepsilon_{l}(q)
-\left(\frac{m_{h}}{m_{l}}-1\right)\varepsilon_{f}\leq
\frac{\varepsilon_{h}(q)}{1-m_{l}/m_{h}}\nonumber\\
 & & \Leftrightarrow\qquad 0\leq\frac{\hbar^{2}}{2m_{l}}
\left(\left(1-\frac{m_{l}}{m_{h}}\right)k_{h}-q\right)^{2}\,.\nonumber
\end{eqnarray}
Moreover, the upper and lower boundary of region II intersect each other
in the $q$-$\omega$-plane at $q=(1-m_{l}/m_{h})k_{h}$ with identical tangent.
Finally, the corresponding contributions to $I_{hl}^{+}(\vec q,\omega)$
in regions I and II are obtained by elementary integration of (\ref{appiHL1})
and given in Eqs.~(\ref{IHL1}),(\ref{IHL2}).

Let us now turn to the lower sign case $I_{hl}^{-}(\vec q,\omega)$. 
The condition 
$\underline{l}_{-}\leq k_{h}$ implies for $b_{-}\geq a^{2}$
($\Leftrightarrow\hbar\omega\geq\varepsilon_{l}(q)$)
\begin{equation}
\hbar q\frac{\hbar k_{h}}{m_{l}}+\varepsilon_{l}(q)
+\left(\frac{m_{h}}{m_{l}}-1\right)\varepsilon_{f}\geq \hbar\omega\,.
\label{ineqHL5}
\end{equation}
In the opposite case $b_{-}\leq a^{2}$ 
($\Leftrightarrow\hbar\omega\leq\varepsilon_{l}(q)$) the inequality
$\underline{l}_{-}\leq k_{h}$ does not lead to any further restriction
on the frequency for $k_{h}\geq a$ 
($\Leftrightarrow(1-m_{l}/m_{h})k_{h}\geq q$), while for $k_{h}\leq a$ 
one finds
\begin{equation}
\hbar\omega\geq-\hbar q\frac{\hbar k_{h}}{m_{l}}+\varepsilon_{l}(q)
+\left(\frac{m_{h}}{m_{l}}-1\right)\varepsilon_{f}\,.
\label{ineqHL6}
\end{equation}
The latter statement is a nontrivial requirement only if
\begin{equation}
q\geq\left(1+\sqrt{m_{l}/m_{h}}\right)k_{h}\,,
\label{ineqHL7}
\end{equation}
which also ensures $(1-m_{l}/m_{h})k_{h}\leq q$ ($\Leftrightarrow k_{h}\leq a$).
The inequalities (\ref{ineqHL5}) and (\ref{ineqHL6})
are necessary and sufficient conditions for $I_{hl}^{-}(\vec q,\omega)$
in (\ref{appiHL1}) to be nonzero. In this case the lower integration bound is 
$\underline{l}_{-}=|a-\sqrt{b_{-}}|=:\underline{k}_{h}^{-}$
and again given explicitly in Eq.~(\ref{lowHL}).

Moreover, straightforward inspection shows that the upper integration
boundary is $\overline{l}_{-}=k_{h}$
provided inequality (\ref{ineqHL6}) ( but not necessarily (\ref{ineqHL7}))
is fulfilled, while otherwise (requiring $q\leq(1-\sqrt{m_{l}/m_{h}})k_{h}$)
we have $\overline{l}_{-}=a+\sqrt{b_{-}}=:\overline{k}_{h}^{-}$ as given in
Eq.~(\ref{highHL}). The corresponding contributions $G_{-}(\dots)$
in Eqn.~(\ref{IHL3}),(\ref{IHL4}) follow again from
elementary integration. The different regions in the $q$-$\omega$-plane
are summarized in table~\ref{tableHL} and illustrated in Fig.~\ref{fig1}.

\subsubsection{$I_{lh}(\vec q,\omega)$}

The contribution $I_{lh}(\vec q,\omega)$ can formally be expressed
by Eq.~(\ref{appiHL1}) via the replacement $h\leftrightarrow l$.
Thus one needs to discuss the condition
\begin{equation}
\Theta\left(k-\left|\frac{q}{2}+\left(1-\frac{m_{h}}{m_{l}}\right)
\frac{k^{2}}{2q}\pm\frac{m_{h}\omega}{\hbar q}\right|\right)=1\,,
\end{equation}
or, equivalently,
\begin{equation}
(k+c)^{2}\geq d_{\pm}\quad\wedge\quad(k-c)^{2}\leq d_{\pm}
\label{ineqcd}
\end{equation}
with
\begin{eqnarray}
c & = & \frac{q}{m_{h}/m_{l}-1}\,,\\
d_{\pm} & = & c^{2}+\frac{2q}{m_{h}/m_{l}-1}
\left(\frac{q}{2}\pm\frac{m_{h}\omega}{\hbar q}\right)\nonumber\\
 & = & \frac{(m_{h}/m_{l})q^{2}}{(m_{h}/m_{l}-1)^{2}}
\pm\frac{2m_{h}\omega/ \hbar}{m_{h}/m_{l}-1}\,.
\end{eqnarray}
Note that, differently from the previous cases, $I_{lh}^{+}(\vec q,\omega)$ 
is not necessarily zero if $I_{lh}^{-}(\vec q,\omega)$ vanishes,
because $1-m_{h}/m_{l}<0$. On the other hand, this inequality ensures
$c\geq 0$, and from the condition $d_{\pm}\geq 0$ we find for the lower case
\begin{equation}
\hbar\omega\leq\frac{1}{m_{h}/m_{l}-1}\varepsilon_{l}(q)\,.
\label{ineqd}
\end{equation}

Similarly to the previous case, the inequalities (\ref{ineqcd}) lead
to the following lower and upper boundaries 
within the interval $[0,k_{l}]$,
\begin{eqnarray}
\underline{l}_{\pm} & = & \min\left\{|c-\sqrt{d_{\pm}}|,k_{l}\right\}\,,\\
\overline{l}_{\pm} & = & \min\left\{c+\sqrt{d_{\pm}},k_{l}\right\}\,,
\end{eqnarray}
with nonzero contributions being possible 
only for $\underline{l}_{\pm}\leq k_{l}$. For the upper sign case this
condition is equivalent to
\begin{equation}
\hbar q\frac{\hbar k_{l}}{m_{h}}-\varepsilon_{h}(q)
-\left(\frac{m_{l}}{m_{h}}-1\right)\varepsilon_{f}\geq \hbar\omega
\label{ineqLH1}
\end{equation}
which can only be fulfilled if 
\begin{equation}
q\leq\left(1+\sqrt{m_{h}/m_{l}}\right)k_{l}\,.
\label{ineqLH2}
\end{equation}
Thus, in the above case, the lower integration bound is 
$\underline{l}_{+}=|c-\sqrt{d_{+}}|=:\underline{k}_{l}^{+}$
and given explicitly in (\ref{lowLH}).
Moreover, an again straightforward discussion shows that the upper
integration bound is given by 
$\overline{l}_{+}=c+\sqrt{d_{+}}=:\overline{k}_{l}^{+}$ (cf. Eq.~(\ref{highLH}))
provided
\begin{equation}
\hbar\omega\leq-\hbar q\frac{\hbar k_{l}}{m_{h}}-\varepsilon_{h}(q)
-\left(\frac{m_{l}}{m_{h}}-1\right)\varepsilon_{f}\,,
\label{ineqLH3}
\end{equation}
which is only possible for
\begin{equation}
q\leq\left(-1+\sqrt{m_{h}/m_{l}}\right)k_{l}\,.
\label{ineqLH4}
\end{equation}
The above inequalities (\ref{ineqLH3}), (\ref{ineqLH4}) define region
II in table~\ref{tableLH} and Fig.~\ref{fig1}, while region I is
defined by (\ref{ineqLH1}),(\ref{ineqLH2}) and the negation of
(\ref{ineqLH3}). Here the upper integration bound is
$\overline{l}_{+}=k_{l}$. The corresponding expressions
for $I_{lh}^{+}(\vec q,\omega)$ in regions I, II are given in
Eqs.~(\ref{ILH1}),(\ref{ILH2}), respectively.

Regarding $I_{lh}^{-}(\vec q,\omega)$, considerations analogous to the
ones for $I_{hl}^{+}(\vec q,\omega)$ show that the 
condition $\underline{l}_{-}\leq k_{l}$ is for $c\geq k_{l}$
($\Leftrightarrow q\geq(m_{h}/m_{l}-1)k_{l}$)
equivalent to
\begin{equation}
\hbar q\frac{\hbar k_{l}}{m_{h}}+\varepsilon_{h}(q)
+\left(\frac{m_{l}}{m_{h}}-1\right)\varepsilon_{f}\geq \hbar\omega
\label{ineqLH5}
\end{equation}
and 
\begin{equation}
\hbar\omega\geq-\hbar q\frac{\hbar k_{l}}{m_{h}}+\varepsilon_{h}(q)
+\left(\frac{m_{l}}{m_{h}}-1\right)\varepsilon_{f}\,,
\label{ineqLH6}
\end{equation}
where latter inequality poses a nontrivial requirement only if
\begin{equation}
q\geq\left(1+\sqrt{m_{h}/m_{l}}\right)k_{h}\,.
\label{ineqLH7}
\end{equation}
The inequalities (\ref{ineqLH5}), (\ref{ineqLH6}), and (\ref{ineqLH7})
define region III in table~\ref{tableLH}. 
Here the lower integration bound is 
$\underline{l}_{-}=|c-\sqrt{d_{-}}|=:\underline{k}_{l}^{-}$
(cf. Eq.~(\ref{lowLH})), and the upper integration bound
turns out to be always $\overline{l}_{-}=k_{l}$. 

For $c\leq k_{l}$, i.e.
\begin{equation}
q\leq\left(m_{h}/m_{l}-1\right)k_{l}\,,
\label{ineqLH8}
\end{equation}
however, there is, similar to the case of $I_{hl}^{+}(\vec q,\omega)$,
another way of fulfilling the condition $\underline{l}_{-}\leq k_{l}$.
The corresponding region IV is defined by the inequalities
(\ref{ineqd}), (\ref{ineqLH8}), and 
\begin{equation}
\hbar\omega\geq\hbar q\frac{\hbar k_{l}}{m_{h}}+\varepsilon_{h}(q)
+\left(\frac{m_{l}}{m_{h}}-1\right)\varepsilon_{f}\,,
\label{ineqLH9}
\end{equation}
and the integration bounds are 
$\underline{l}_{-}=\underline{k}_{l}^{-}$,
$\overline{l}_{-}=c+\sqrt{d_{-}}=:\overline{k}_{l}^{-}$,

We note that the inequality (\ref{ineqLH5}) generally implies the 
fulfillment of (\ref{ineqd}) since
\begin{eqnarray}
 & &\hbar q\frac{\hbar k_{l}}{m_{h}}+\varepsilon_{h}(q)
+\left(\frac{m_{l}}{m_{h}}-1\right)\varepsilon_{f}\leq
\frac{\varepsilon_{l}(q)}{m_{h}/m_{l}-1}\nonumber\\
 & & \Leftrightarrow\qquad 0\leq\frac{\hbar^{2}}{2m_{h}}
\left(\left(\frac{m_{h}}{m_{l}}-1\right)k_{l}-q\right)^{2}\,.\nonumber
\end{eqnarray}
Moreover, similarly as in the case of 
$I_{hl}^{+}(\vec q,\omega)$, 
the upper and the lower boundary of region IV intersect each other
at $q=(m_{h}/m_{l}-1)k_{l}$ with identical tangent.


\begin{thebibliography}{}
\bibitem{notechi}
The contributions $\chi_{hh}(\vec q,\omega)$, $\chi_{ll}(\vec q,\omega)$
correspond to the single-particle excitations among heavy holes only
and light holes only, respectively. However, such an interpretation is,
strictly speaking, not possible for $\chi_{hl}(\vec q,\omega)$ and 
$\chi_{lh}(\vec q,\omega)$ as separate quantities. 
In particular, these terms are not identical to
the contributions to the r.h.s of Eq.~(\ref{chi01}) with 
$|\lambda_{1}|=3/2$, $|\lambda_{2}|=1/2$ and 
$|\lambda_{1}|=1/2$, $|\lambda_{2}|=3/2$, respectively. This is due to the
fact that by deriving Eq.~(\ref{chi02}) from Eq.~(\ref{chi01}) 
via a standard procedure\cite{Giuliani05,Mahan00,Bruus04}, the role of
the band indices is interchanged in a part of the expression. As a
result, only the sum $\chi_{hl}(\vec q,\omega)+\chi_{lh}(\vec q,\omega)$
is identical to the sum of all terms in Eq.~(\ref{chi01}) with
$|\lambda_{1}|\neq|\lambda_{2}|$ and allows for an interpretation
in terms of single-particle excitations\cite{Giuliani05,Mahan00,Bruus04}.
\bibitem{notecorr1}
The expression (\ref{static}) corrects some typographical errors in Eq.~(4) of 
Ref.~\cite{Schliemann10} which do not affect any other part of the
content there.
\bibitem{notereal}
In order to efficiently display the zeros of the real part of the
dielectric function, the modulus  $|\varepsilon^{RPA}(\vec q,\omega)|$
(not $\varepsilon^{RPA}(\vec q,\omega)$ itself) is plotted in 
Figs.~\ref{fig2},\ref{fig5}.
\bibitem{notecorr2}
The result~(\ref{high}) corrects in detail an expression given previously
in Ref.~\cite{Schliemann10}. Corresponding changes apply to the coefficients
$u$ and $w$ in Eqs.~(\ref{u}) and (\ref{w}), respectively, compared to
 Ref.~\cite{Schliemann10}. However, these corrections do not change
the qualitative properties of the high-frequency expansion of the
dielectric function. In particular, the salient feature of terms being of
low order in the wave vector and occurring in any order in the inverse frequency
remains unaltered.
\end{thebibliography}
\end{document}